\documentclass[
prl,preprint,
superscriptaddress,
showpacs,
amsmath,amssymb,
aps
]{revtex4-1}

\usepackage{graphicx}
\usepackage{amsmath}
\usepackage[version=3]{mhchem}
\usepackage[usenames]{color}

\usepackage{braket}
\usepackage{color}

\usepackage{amsmath}

\usepackage{braket}

\usepackage[colorlinks=true, linkcolor=blue, citecolor=blue, urlcolor=blue,pdftex]{hyperref}
\definecolor{newcolor}{rgb}{0.9,0,0.1}

\newcommand{\figref}[1]{Fig.~\ref{#1}}
\newcommand{\unit}[2]{$#1\,\text{#2}$}
\newcommand{\equnit}[3]{$#1=#2\,\text{#3}$}

\begin{document}

\title{Electrically driven photon emission from individual atomic defects in monolayer \ce{WS2}}



\author{Bruno Schuler}
\email[]{bschuler@lbl.gov}

\author{Katherine A. Cochrane}
\affiliation{Molecular Foundry, Lawrence Berkeley National Laboratory, California 94720, USA}

\author{Christoph Kastl}
\affiliation{Molecular Foundry, Lawrence Berkeley National Laboratory, California 94720, USA}
\affiliation{Walter-Schottky-Institut and Physik-Department, Technical University of Munich, Garching 85748, Germany}

\author{Ed Barnard}

\author{Ed Wong}
\affiliation{Molecular Foundry, Lawrence Berkeley National Laboratory, California 94720, USA}

\author{Nicholas Borys}
\affiliation{Molecular Foundry, Lawrence Berkeley National Laboratory, California 94720, USA}
\affiliation{Department of Physics, Montana State University, Bozeman, Montana 59717, USA}

\author{Adam M. Schwartzberg}

\author{D. Frank Ogletree}
\affiliation{Molecular Foundry, Lawrence Berkeley National Laboratory, California 94720, USA}

\author{F. Javier Garc\'{i}a de Abajo}
\email[]{javier.garciadeabajo@icfo.eu}
\affiliation{ICFO-Institut de Ciencies Fotoniques, The Barcelona Institute of Science and Technology, 08860 Castelldefels (Barcelona), Spain}
\affiliation{ICREA-Instituci\'o Catalana de Recerca i Estudis Avan\c{c}ats, 08010 Barcelona, Spain}

\author{Alexander Weber-Bargioni}
\email[]{afweber-bargioni@lbl.gov}
\affiliation{Molecular Foundry, Lawrence Berkeley National Laboratory, California 94720, USA}



\begin{abstract}
Optical quantum emitters are a key component of quantum devices for metrology and information processing. In particular, atomic defects in 2D materials can operate as optical quantum emitters that overcome current limitations of conventional bulk emitters, such as yielding a high single-photon generation rate and offering surface accessibility for excitation and photon extraction. Here we demonstrate electrically stimulated photon emission from individual point defects in a 2D material. Specifically, by bringing a metallic tip into close proximity to a discrete defect state in the band gap of \ce{WS2}, we induce inelastic tip-to-defect electron tunneling with an excess of transition energy carried by the emitted photons. We gain atomic spatial control over the emission through the position of the tip, while the spectral characteristics are highly customizable by varying the applied tip-sample voltage. Atomically resolved emission maps of individual sulfur vacancies and chromium substituent defects are in excellent agreement with the electron density of their respective defect orbitals as imaged \textit{via} conventional elastic scanning tunneling microscopy. Inelastic charge-carrier injection into localized defect states of 2D materials thus provides a powerful platform for electrically driven, broadly tunable, atomic-scale single-photon sources.
\end{abstract}

\maketitle

Solid-state quantum emitters are promising building blocks for quantum technologies such as high-precision sensing and secure communications~\cite{aharonovich2016solid}. In particular, atom-like emitters embedded in a host crystal (e.g., color centers in diamond~\cite{gruber1997scanning,sukachev2017silicon} and silicon carbide (SiC)~\cite{castelletto2014silicon}) combine several appealing properties; namely, spin-selective optical transitions~\cite{gruber1997scanning}, room temperature stability~\cite{maurer2012room}, exceptionally long coherence times~\cite{maurer2012room}, and potential for scalability~\cite{koehl2015designing}. However, the quest for the ideal on-demand single-photon emitter is still ongoing to address several materials science and engineering challenges. This includes controlling the mesoscopic environment to avert variability between emitters; identifying and eliminating decoherence channels; achieving precision spatial placement; and developing high-fidelity, scalable pumping schemes that are amenable to on-chip integration, such as electrically driven optical emission~\cite{aharonovich2016solid}.\\

Two-dimensional (2D) materials offer several key advantages over their bulk counterparts as a platform for quantum photonics~\cite{aharonovich2016solid,mak2016photonics}, such as synthetic flexibility~\cite{das2015beyond}, higher photon extraction efficiency~\cite{aharonovich2016solid}, and tunability through external fields and choice of substrate~\cite{novoselov20162d}. They also provide better control of emitter placement~\cite{kern2016nanoscale} and integration with photonic~\cite{liu2015strong,wu2015monolayer} and plasmonic~\cite{butun2015enhanced} nanocavities. In particular, semiconducting transition metal dichalcogenides (TMDs) feature unique valleytronic~\cite{xu2014spin} and magneto-optical effects~\cite{wang2017valley} with great potential for customizing the photon emission. Recently, single-photon emitters have been reported in hBN~\cite{tran2016quantum}, \ce{WSe2}~\cite{chakraborty2015voltage,srivastava2015optically,he2015single,koperski2015single,kumar2015strain}, and \ce{MoSe2}~\cite{chakraborty2016localized}, while electrically driven quantum light emitting diodes (QLEDs) have been demonstrated with \ce{WSe2} as the optically active layer in a vertical tunneling junction ~\cite{Palacios-Berraquero2016,clark2016single}. However, the identification of the actual atomic origin of the emission (i.e., which defect drives what specific emission) still remains unresolved, thus precluding the desirable control over atomic-scale placement, electrical injection, and ultimately, the resulting photon generation. \\

\begin{figure*}[]
\includegraphics[width=0.7\textwidth]{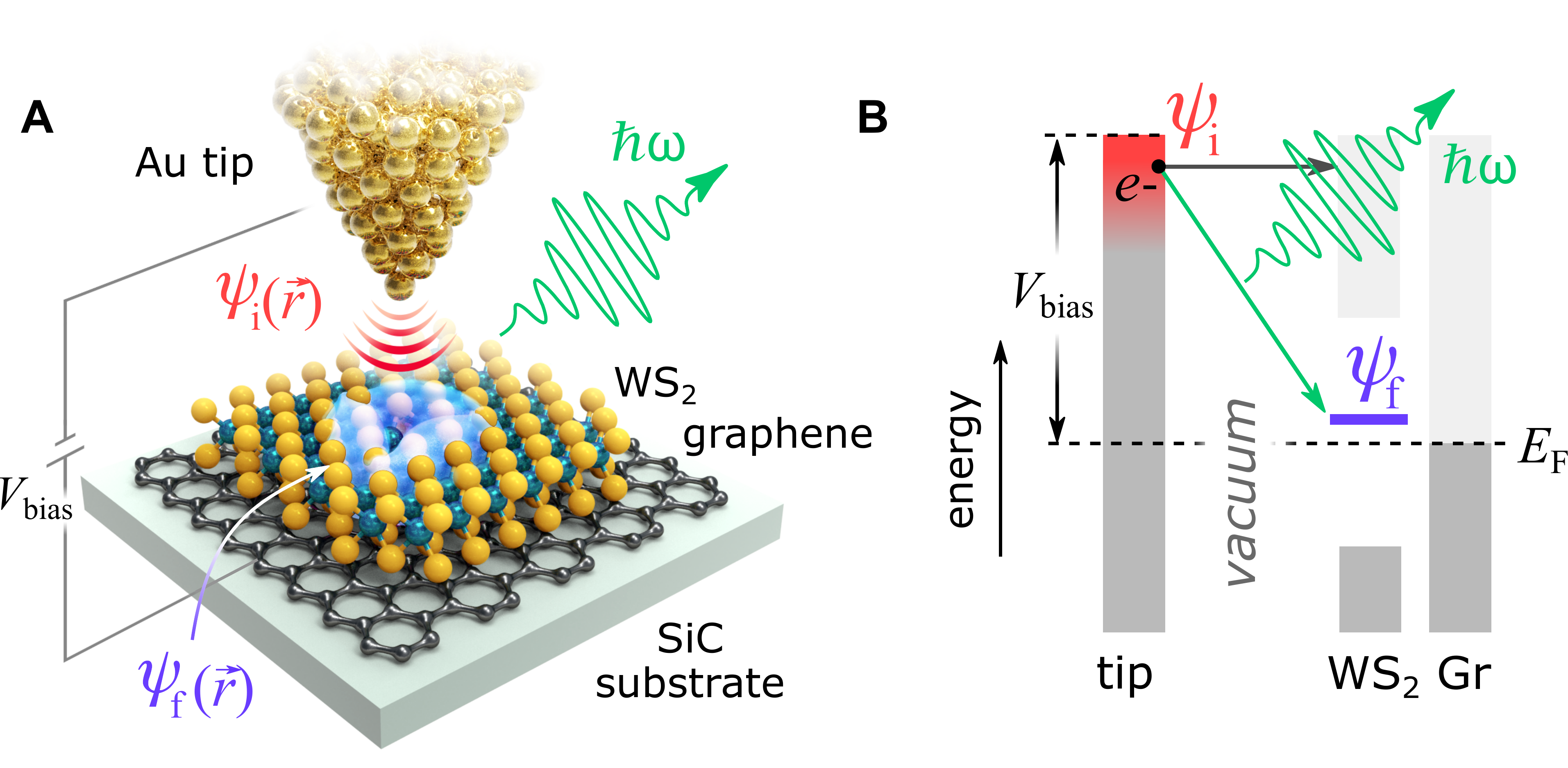}
\caption{\label{fig:schematic}
\textbf{Scheme of experimental configuration (A) and electron energy levels (B) involved in electrically driven photon emission from an atomic point defect.} An electron from a gold tip (initial state $\psi_\text{i}$ in red) tunnels inelastically into an atomic defect (top sulfur vacancy) of monolayer \ce{WS2} (final state $\psi_\text{f}$ in purple). The excess of transition energy is released into the emission of one photon, mediated by intermediate plasmon states of the tip cavity. The range of photon emission energies is controlled through the tip-sample bias voltage. The 2D material is supported on a graphene layer (electric contact) on top of SiC. The dark and light gray boxes in B represent filled and empty states, respectively.}
\end{figure*}

Here, we demonstrate electrically stimulated photon emission from individual atomic emitters in a 2D material. Specifically, we observe single-defect luminescence in monolayer \ce{WS2} driven by electron tunneling from a metallic tip with precise atomic control over the source of the emission (\figref{fig:schematic}A). Electrons injected from the continuum of metallic tip states into discrete defect states of the 2D material generate broad optical emission spectra, with associated photon energies corresponding to the difference between initial (tip) and final (defect) electron energies in the tunneling process (\figref{fig:schematic}B). The metal tip acts as a plasmonic cavity that assists the coupling between the inelastic tunneling current and the emitted light far-field. This allows us to control the emission spectrum through the applied tip-sample bias voltage, while spatial control is enabled by the proportionality between the emission rate and the electron density in the defect orbital right under the tip position. We draw these conclusions by studying deliberately introduced sulfur vacancies and native chromium substituent defects in monolayer \ce{WS2}, for which we present spatial maps of the photon emission rate as a function of tip position, revealing the defect orbitals with ultimate atomic resolution on par with that obtained using conventional elastic scanning tunneling microscopy.\\

We use the Au-coated tip of a scanning tunneling microscope (STM) to inject charge carriers and mediate the coupling between optical near- and far-fields {\it via} tip plasmon modes. This is the so-called STM luminescence~\cite{kuhnke2017atomic} (STML), which has been successful in studying metallic surfaces~\cite{berndt1991inelastic} and molecular systems~\cite{wu2008intramolecular}, beating the light diffraction limit by more than two orders of magnitude due to the extreme localization of initial and final electron states in the tunneling process. In this context, electrofluorescence from single molecules has been recently established through self-decoupling~\cite{merino2015exciton} or by introducing ultrathin insulating layers~\cite{yu2016tunneling,imada2016real,doppagne2018electrofluorochromism} between the molecule and a noble metal substrate. Additionally, STML has enabled vibronic spectroscopy with submolecular resolution~\cite{doppagne2017vibronic,doppagne2018electrofluorochromism}, imaging  molecular orbitals through photon emission maps~\cite{lutz2013molecular,yu2018visualization}, studying charge and exciton dynamics~\cite{merino2015exciton,roslawska2018single,merino2018bimodal}, and initiating spin-selective optical transitions~\cite{kimura2019selective}. 
STML on ITO-supported \ce{MoSe2} has also been reported and  attributed to radiative decay of the A exciton in this material~\cite{pommier2019scanning}.\\

Plasmonic noble metal substrates, which are commonly employed in STML, are not a viable option to investigate TMDs because strong hybridization quenches the intrinsic optical emission~\cite{krane2016electronic}. Instead, we use epitaxial graphene grown on silicon carbide (SiC) as a substrate, which has been shown to preserve the native TMD band structure~\cite{forti2017electronic,Kastl2017cvd}.
While electroluminescence from graphene has been previously observed~\cite{beams2014electroluminescence}, we find conclusive evidence to assign the optical emission in our TMD/graphene heterostructure to the electronic states of the TMD alone.\\

\begin{figure*}[]
\includegraphics[width=0.7\textwidth]{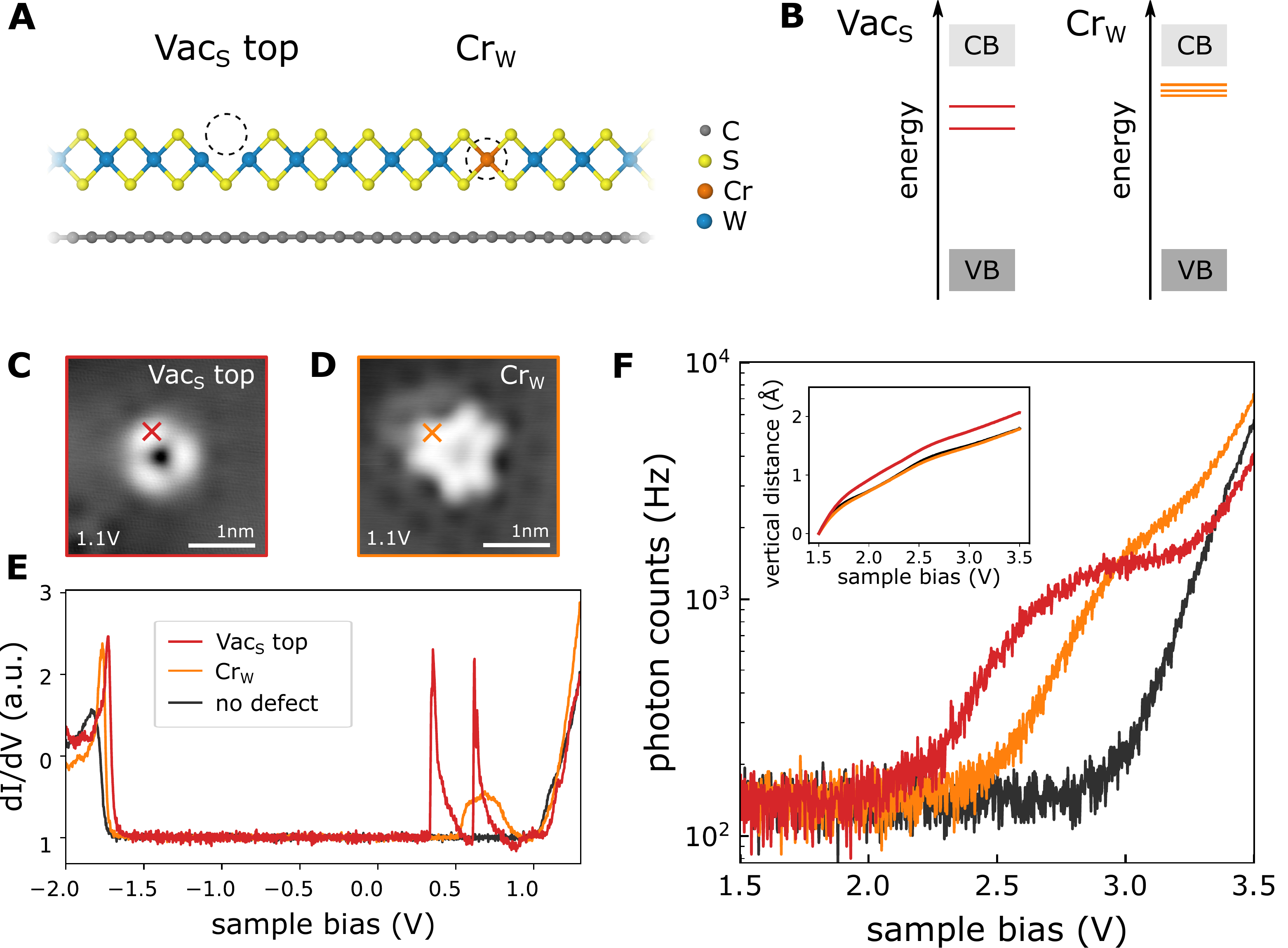}
\caption{\label{fig:EmissionVsBias}
\textbf{Tunneling bias dependence of photon emission.} (\textbf{A}) Side view of atomic layers and the two types of defects studied here in \ce{WS2}/graphene: a top sulfur vacancy (Vac$_\text{S}$ top) and a chromium substituting tungsten (Cr$_\text{W}$). (\textbf{B}) Energy level diagrams showing the two unoccupied Vac$_\text{S}$ defect states (red) and three unoccupied Cr$_\text{W}$ defect states (orange), lying in the gap between valence and conduction bands (VB, CB). (\textbf{C,D}) STM topography (constant current \equnit{I}{20}{pA} and bias \equnit{V}{1.1}{V}) of Vac$_\text{S}$ top (C) and Cr$_\text{W}$ (D). (\textbf{E}) We probe the local density of electron states (LDOS) through $dI/dV$ spectroscopy on Vac$_\text{S}$ top (red) and Cr$_\text{W}$ (orange) defects at positions indicated by crosses in C, D, compared with a position far from any defect (gray). (\textbf{F}) Spectrally integrated photon emission induced by inelastic electron tunneling at constant current \equnit{I}{10}{nA}. Measurements in E and F were recorded at the same defect sites and with the same tip.
}
\end{figure*}

We prepare our samples by growing monolayer \ce{WS2} islands on epitaxial graphene on SiC using chemical vapor deposition~\cite{Kastl2017cvd}. The as-grown sample contains several substitutional atomic defects, such as chromium (Cr$_\text{W}$) and molybdenum (Mo$_\text{W}$) replacing tungsten, as well as oxygen substituting sulfur (O$_\text{S}$)~\cite{schuler2019overview,barja2018identifying}. Then, we selectively generate sulfur vacancies (Vac$_\text{S}$) by high temperature annealing in vacuum~\cite{schuler2018large}. Both Cr$_\text{W}$ and Vac$_\text{S}$ defects exhibit unoccupied electronic states placed a few hundred millielectronvolts below the \ce{WS2} conduction band edge~\cite{schuler2018large,schuler2019overview} (\figref{fig:EmissionVsBias}B). These gap defect states give rise to pronounced resonances in the differential conductance ($dI/dV$) spectra measured in scanning tunneling spectroscopy (STS), which roughly yields the local density of electron state (LDOS), as shown in \figref{fig:EmissionVsBias}E. Their characteristic electronic spectrum results from a combination of crystal-field splitting, spin-orbit coupling, and electron-phonon interaction~\cite{schuler2019overview}.\\

At tunneling voltages exceeding \unit{1.5}{V}, we observe electron-induced photon emission on the Cr$_\text{W}$ and Vac$_\text{S}$ defects, as well as on defect-free \ce{WS2} locations. The emission rate strongly depends on the applied bias and position (e.g., on or off a defect, as well as different regions within the defect) where STML spectra are acquired at constant current, as shown in \figref{fig:EmissionVsBias}F. 
Importantly, we find a clear correlation between the bias onset for photon emission and the energy of the lowest unoccupied states observed in $dI/dV$. In particular, the difference between the bias onset of Vac$_\text{S}$ (\unit{2.2}{V}) and Cr$_\text{W}$ (\unit{2.4}{V}) is \unit{0.2}{V}, which agrees with the energy difference of their respective defect states. Similarly, we observe STML at negative bias polarities (hole injection), for which the emission onset scales with the energy of the highest occupied state (see Fig.~S7). The emission scales linearly with the current, thus suggesting a single-electron process.\\

\begin{figure*}[]
\includegraphics[width=0.7\textwidth]{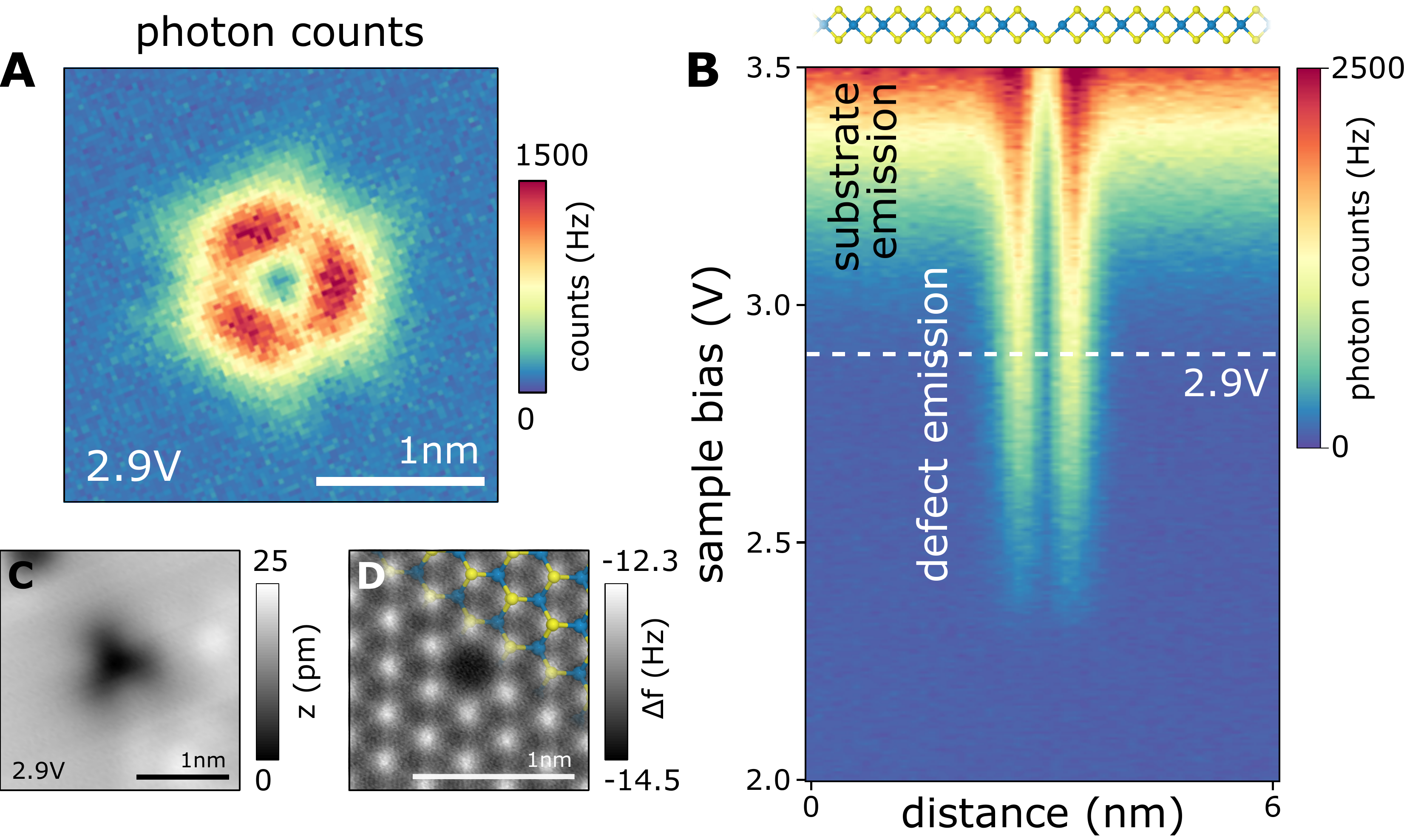}
\caption{\label{fig:EmissionVsPosition}
\textbf{Photon-emission spatial imaging of a single sulfur vacancy.} (\textbf{A}) Spectrally integrated photon map of a sulfur top vacancy using \equnit{I}{20}{nA} constant current and \equnit{V}{2.9}{V} bias. The photon map agrees well with the STM image of the in-gap defect orbital (see \figref{fig:EmissionVsBias}C). (\textbf{B}) Spectrally integrated photon emission across Vac$_\text{S}$ top as a function of tunneling bias. The dashed white line indicates the bias at which the map in A was acquired. The defect exhibits strongly localized photon emission at a tunneling bias significantly lower than the onset of substrate emission.
(\textbf{C}) STM topography (\equnit{I}{20}{nA}) of Vac$_\text{S}$ top simultaneously acquired during the photon map shown in A. (\textbf{D}) CO-tip nc-AFM image of a Vac$_\text{S}$ top. 
}
\end{figure*}

The extremely localized excitation by tunneling electron injection allows us to record atomically resolved photon maps. In \figref{fig:EmissionVsPosition}A we show the spectrally integrated photon emission as a function of lateral tip position over a single sulfur vacancy. The subnanometer-resolved photon maps acquired at high biases closely resemble the STM image of the in-gap defect orbital. The emission does not correlate with the simultaneously acquired STM topography at high bias (\figref{fig:EmissionVsPosition}C), therefore excluding any significant effect of the slightly varying gap distance on the spatial variation of the emission. Laterally, the defect emission is closely confined within $\sim$\unit{1}{nm}, concurrent with the electronic orbital dimensions (see \figref{fig:EmissionVsPosition}B).\\

The close resemblance between STM and STML maps is further supported by theory. Indeed, photon emission in STML is mediated by the transition dipole ${\bf p}$ associated with the tunneling electron, acting as a radiation source~\cite{JMA1990} and giving rise to an emission yield $\propto|{\bf p}|^2$, which is in turn proportional to the elastic tunneling current measured in STM (see details in the Supplemental Information, SI). This current alone bears a dependence on the sampled final state, and therefore, both elastic and inelastic tunneling rates are proportional to the electron density of states at the tip position, given by the defect orbital density $|\psi_\text{f}|^2$. Similar to the enhanced emission rate of an excited atom in a resonant cavity defined by the Purcell factor $P(\omega)$~\cite{P1946}, the spectrally resolved STML rate is directly proportional to $P(\omega)$, which for metallic tips is dominated by light-plasmon coupling and can be observed through optical spectroscopy.\\

\begin{figure*}[]
\includegraphics[width=0.8\textwidth]{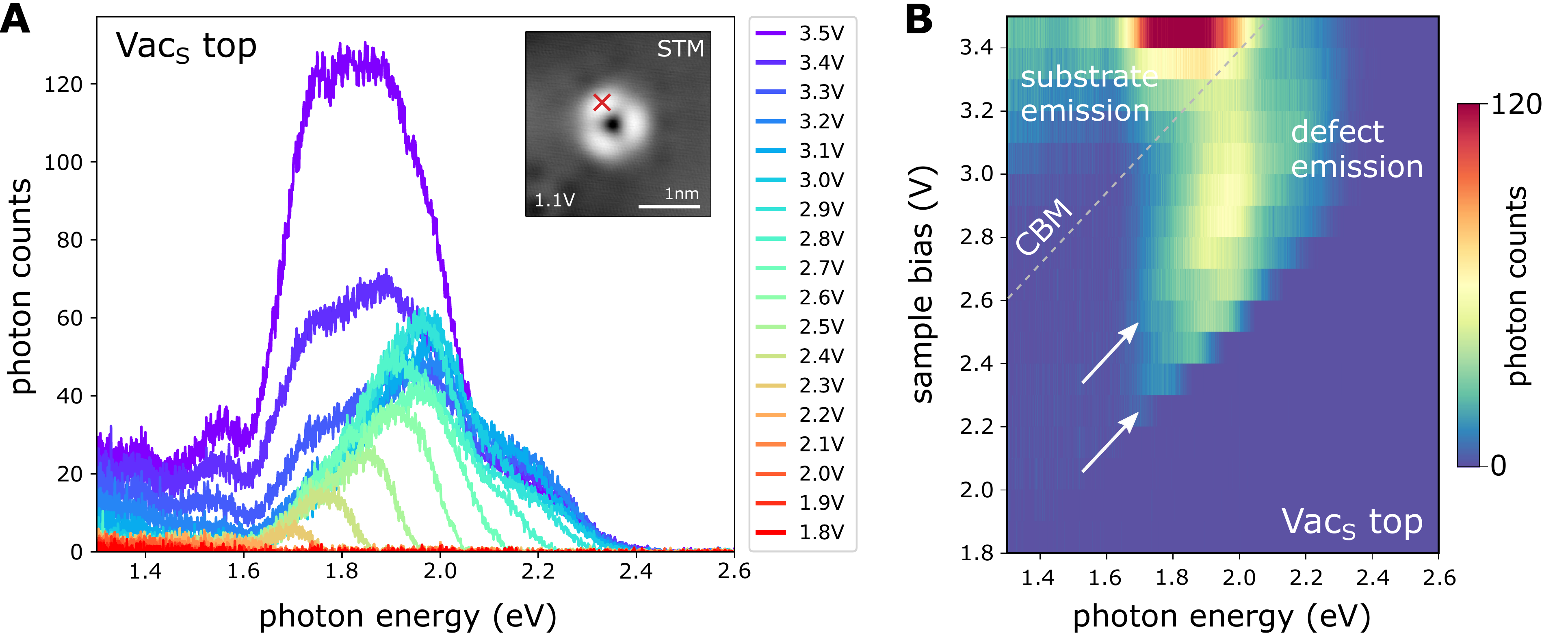}
\caption{\label{fig:OpticalSpectrum}
\textbf{Spectrally resolved photon emission on a single sulfur vacancy.} (\textbf{A}) Photon emission spectrum on Vac$_\text{S}$ top at different tunneling biases. The spectral width and the minimal spectral profile variation with lateral tip position suggest a plasmon-mediated emission process. (\textbf{B}) 2D representation of the data in A. Oblique arrows indicate the calculated maximum photon energy from transitions involving the two discrete Vac$_\text{S}$ defect states. The dashed line marks the onset of substrate emission corresponding to tunneling to its conduction band minimum (CBM). 
}
\end{figure*}

Spectrally resolved STML measurements on the sulfur vacancy defect shown in \figref{fig:OpticalSpectrum} reveal a broadband photon emission. The optical emission band is centered around \unit{1.9}{eV}, in good agreement with the expected plasmon frequency range of our Au tip (i.e., where the Purcell enhancement $P(\omega)$ is maximum, see SI). We can identify two distinct regions in the plot of the emission intensity as a function of photon energy and tunneling bias (\figref{fig:OpticalSpectrum}B): the high-bias, low-photon-energy corner (top left) is associated with substrate emission from \ce{WS2} (i.e., tunneling to substrate states), while the distinct emission band at lower bias is associated exclusively with the defect. For Vac$_\text{S}$ two emission steps (marked by white arrows in \figref{fig:OpticalSpectrum}B) are observed, which follow a linear relation between applied bias and photon energy. Each of these steps corresponds to the opening of a new radiative decay channel, which we attribute to a transition from tip states at or below its Fermi level into the two unoccupied sulfur vacancy states. Emission of a single photon associated with a single-electron transition is the dominant lowest-order radiative process. Accordingly, the highest allowed photon energy is given by the difference between the tip Fermi energy and the defect state energy.\\

For our Au coated tungsten tips, we detect $\sim10^{-7}$ far-field photons per electron. Accounting for all setup related losses, the intrinsic quantum efficiency of the radiative tunneling process is estimated as $Y\sim10^{-4}$ photons per electron. It is interesting to compare this number with the analytical result (see SI) $Y\sim(3\pi\alpha/2)(d/\lambda)^2(\Delta\omega/\omega)P(\omega)$, where $\omega$, $\lambda$, and $\Delta\omega$ are the photon frequency, wavelength, and bandwidth, $d$ is the tip-sample distance, and $\alpha\approx1/137$ is the fine-structure constant. The Purcell factor $P$, which scales linearly with the density of available optical states dominated by plasmon modes in the tip cavity \cite{paper156}, can be estimated by invoking electromagnetic reciprocity, asserting that the emission intensity is strictly proportional to the enhancement in the near-field intensity under external illumination at the position of the emitting dipole. The latter depends on the precise tip-sample morphology but is rather insensitive to the lateral position of the tip in the nonmetallic SiC substrate under consideration and reaches plasmon-enhanced values of $P\sim10^4$  (see SI). Using parameters corresponding to our experimental conditions, we then predict a photon yield per tunneled electron $Y\sim10^{-4}$, in good agreement with the experimental estimate.

Ultimately, the photon generation rate is limited by Coulomb blockade, which prevents tunneling of subsequent electrons into the defect state before the previous electron drains to the graphene substrate. The emission is thus intrinsically consisting of single photons. From the linewidth of the vacancy state, we estimate the transient charging time of the defect to be $<$\unit{100}{fs}, hence indicating that a \unit{10^9}{Hz} single-photon generation rate should be attainable. This number may be easily increased using optimized plasmonic or optical cavities to increase the Purcell factor and drive the emission along a desired channel (e.g., a waveguide).\\

In conclusion, we demonstrate electron-stimulated photon emission from individual, atomically-defined defects in a 2D semiconductor. Atomically resolved luminescence maps from deliberately introduced sulfur vacancy defects and native chromium substituents reveal that the in-gap defect orbitals are the final states of the optical transition. Electrons that tunnel inelastically from a continuum of tip states into discrete defect states can convey their excess transition energy into plasmonic excitations in the tip nanocavity, mediating their coupling to propagating photons. The widely tunable optical emission generated by charge carrier injection into localized defect states in a 2D material is a powerful platform for electrically driven single-photon emission. The single-defect / single-electron / single-photon regime allows us to avoid averaging and environmental decoherence effects. Monochromatic electron injection, for instance using a superconducting electrode and charge state control of the defect by gating, could eventually enable on-demand spin-polarized single-photon emission in a solid-state tunneling device.

\bibliographystyle{science}
\bibliography{STML,refs}

\section*{Acknowledgments}
This work was performed at the Molecular Foundry supported by the Office of Science, Office of Basic Energy Sciences, of the U.S. Department of Energy under Contract No. DE-AC02-05CH11231. B.S. appreciates support from the Swiss National Science Foundation under project number P2SKP2\_171770. 
C.K. gratefully acknowledges support by the Bavaria California Technology Center (BaCaTeC) and the International Graduate School of Science and Engineering (IGSSE) via project ``CommOnChip''. A.W.-B. was supported by the U.S. Department of Energy Early Career Award.
J.G.A. acknowledges support from the Spanish MINECO (MAT2017-88492-R and SEV2015-0522), the European Research Council (Advanced Grant 789104-eNANO), the Catalan CERCA Program, and Fundaci{\'o} Privada Cellex.

\clearpage

\end{document}


\title{Supplementary Information:\texorpdfstring{\\}{} Electrically driven photon emission from individual atomic defects in monolayer \ce{WS2}}

\author{Bruno Schuler}
\email[]{bschuler@lbl.gov}

\author{Katherine A. Cochrane}
\affiliation{Molecular Foundry, Lawrence Berkeley National Laboratory, California 94720, USA}

\author{Christoph Kastl}
\affiliation{Molecular Foundry, Lawrence Berkeley National Laboratory, California 94720, USA}
\affiliation{Walter-Schottky-Institut and Physik-Department, Technical University of Munich, Garching 85748, Germany}

\author{Ed Barnard}

\author{Ed Wong}
\affiliation{Molecular Foundry, Lawrence Berkeley National Laboratory, California 94720, USA}

\author{Nicholas Borys}
\affiliation{Molecular Foundry, Lawrence Berkeley National Laboratory, California 94720, USA}
\affiliation{Department of Physics, Montana State University, Bozeman, Montana 59717, USA}

\author{Adam M. Schwartzberg}

\author{D. Frank Ogletree}
\affiliation{Molecular Foundry, Lawrence Berkeley National Laboratory, California 94720, USA}

\author{F. Javier Garc\'{i}a de Abajo}
\email[]{javier.garciadeabajo@icfo.eu}
\affiliation{ICFO-Institut de Ciencies Fotoniques, The Barcelona Institute of Science and Technology, 08860 Castelldefels (Barcelona), Spain}
\affiliation{ICREA-Instituci\'o Catalana de Recerca i Estudis Avan\c{c}ats, 08010 Barcelona, Spain}

\author{Alexander Weber-Bargioni}
\email[]{afweber-bargioni@lbl.gov}
\affiliation{Molecular Foundry, Lawrence Berkeley National Laboratory, California 94720, USA}


\begin{abstract}
In this document we derive an analytical expression for the photon emission probability associated with inelastic tunneling in STM and a simple estimate of its ratio relative to the elastic current. We further provide details on sample growth and preparation, as well as the STM luminescence dependence on the type of \ce{WS2} defect, bias polarity, tunneling current, and tip variability.
\end{abstract}

\maketitle 

\tableofcontents

\section{Theoretical modelling}
An expression for the electron intensity associated with elastic tunneling was first obtained by Bardeen \cite{B1961} in terms of the quantum current evaluated in the gap region between two conductors, and later specialized to the tip-near-a-surface geometry that characterized STM experiments \cite{TH1985}. Inelastic tunneling assisted by light emission was eventually observed in the tip-substrate configuration \cite{GRC1988,BGJ91}, motivating further theoretical developments needed to model light emission in STM \cite{JMA1990,J98,AAB00,HAA01}. Further experiments demonstrated the crucial role of plasmons in nanostructured metal samples \cite{NEF02,MSA02}, causing a dramatic dependence on the detailed tip-sample morphology, which is generally unknown.\\

Previous observations of light emission by electron tunneling in heterostructures \cite{LM1976} and their many-body theoretical analysis \cite{PB92} had already indicated that the photon emission yield was only a small fraction of the tunneling current ($<10^{-3}$), so it could be described within first-order perturbation theory, in a way analogous to its elastic counterpart \cite{B1961}. To first-order, light emission in STM involves one quantum of excitation present in the system at any given time (i.e., the electron transition from the initial to the final state, which can possibly couple to an intermediate excitation in the medium such as a plasmon, followed by subsequent decay into an emitted photon). Under these conditions, a full quantum treatment of both the radiation and the generally lossy tip-sample system is fully equivalent to a semi-classical formulation in which the initial and final electronic states (assuming a one-electron picture with wave functions $\psi_{\rm i}(\rb)$ and $\psi_{\rm f}(\rb)$, and energies $\hbar\varepsilon_{\rm i}$ and $\hbar\varepsilon_{\rm f}$, respectively) define an inelastic tunneling current $\jb(\rb,t)=\jb(\rb)\ee^{-\ii\omega t}+\jb^*(\rb)\ee^{\ii\omega t}$ with
\begin{align}
\jb(\rb)=\frac{-\ii e\hbar}{2\me}\left[\psi_{\rm i}(\rb)\nabla\psi^*_{\rm f}(\rb)-\psi^*_{\rm f}(\rb)\nabla\psi_{\rm i}(\rb)\right],
\label{jb}
\end{align}
which we can treat as a classical source at the transition frequency $\omega=\varepsilon_{\rm i}-\varepsilon_{\rm f}$ \cite{GRC1988}. We only need to describe $\ee^{-\ii\omega t}$ components and then use causality to obtain the full time dependence by taking twice the real part of the calculated complex field amplitudes. Dropping the overall $\ee^{-\ii\omega t}$ factor for simplicity, we can write the electric field produced by the above current in terms of the electromagnetic $3\times3$ Green tensor $\mathcal{G}(\rb,\rb',\omega)$ as
\begin{align}
\Eb(\rb)=\frac{\ii}{\omega}\int d^3\rb'\,\mathcal{G}(\rb,\rb',\omega)\cdot\jb(\rb').
\label{EGj}
\end{align}
We further assume a local dielectric description of the system, which allows us to characterize it through a permittivity $\epsilon(\rb,\omega)$ given by the frequency-dependent permittivity of the material present at each position $\rb$ (e.g., $\epsilon=1$ in vacuum). The Green tensor then satisfies the relation \cite{J99}
\begin{align}
\nabla\times\nabla\times\mathcal{G}(\rb,\rb',\omega)&-k^2\epsilon(\rb,\omega)\mathcal{G}(\rb,\rb',\omega) \nonumber\\
&=4\pi k^2\delta(\rb-\rb')\mathcal{I}_3,
\nonumber
\end{align}
where $k=\omega/c$ is the free-space light wave vector and $\mathcal{I}_3$ is the $3\times3$ unit matrix.\\

As an illustration of the generality of this formalism, its application to transitions between the states of a fast electron in the beam of a transmission electron microscope readily leads to a widely used expression for the electron energy-loss probability \cite{paper149}. This approach is also useful to study the decay of excited atoms in the presence of arbitrarily shaped structures \cite{paper053}, where the small size of the associated current distribution $\jb$ compared with the light wavelength allows us to condense it into a transition dipole. For STM-induced photon emission, such transition-dipole approach also constitutes a reasonable approximation \cite{JMA1990}.

\subsection{Emission and decay from a dipole}

In the study of both atomic decay and STM-induced light emission we can generally exploit the fact that the extension of the involved electronic states is small compared with the light wavelength, thus reducing Eq.\ (\ref{EGj}) to
\begin{align}
\Eb(\rb)\approx\mathcal{G}(\rb,\rb_0,\omega)\cdot\pb,
\label{EGp}
\end{align}
where
\begin{align}
\pb=\frac{\ii}{\omega}\int d^3\rb\,\jb(\rb)=\frac{-e\hbar}{\me\omega}\int d^3\rb\,\psi^*_{\rm f}(\rb)\nabla\psi_{\rm i}(\rb)
\label{pb}
\end{align}
is the transition dipole moment (the rightmost expression results from inserting Eq.\ (\ref{jb}) into the integral and integrating one of the terms by parts) and $\rb_0$ denotes to the position of the dipole (e.g., the atom position). Incidentally, when the initial and final electron states are solutions of the Schr\"odinger equation with the same potential, the above expression for the transition dipole reduces to $\pb=-e\int d^3\rb\,\psi^*_{\rm f}(\rb)\psi_{\rm i}(\rb)\,\rb$, although this expression might not be applicable in STM, where the two states can even have different effective masses. In particular, for an excited atom in vacuum (i.e., taking $\mathcal{G}(\rb,\rb_0,\omega)=\mathcal{G}_0(\rb,\rb_0,\omega)=(k^2\mathcal{I}_3+\nabla\otimes\nabla)\ee^{\ii k|\rb-\rb'|}/|\rb-\rb'|$) \cite{J99}, calculating the emitted power from the integral of the outgoing Poynting vector over a distant sphere centered at the atom, and dividing the result by the photon energy $\hbar\omega$, we readily find the photon emission rate (i.e., the atom decay rate, since there are no other decay mechanisms in this configuration) to be $\Gamma_0=4k^3|\pb|^2/3\hbar$, which agrees with the expression obtained from a more tedious procedure involving the quantization of the electromagnetic field \cite{L1983}.\\

The present formalism has also been extensively used to rigorously obtain the decay and emission rates of excited atoms near surfaces and nanostructures \cite{paper053,NH06}. For example, for $\rb$ and $\rb'$ near a planar surface, we can separate $\mathcal{G}=\mathcal{G}_0+\mathcal{G}_{\rm ref}$ as the sum of a free-space component $\mathcal{G}_0$ and a surface-reflection contribution $\mathcal{G}_{\rm ref}$, both of which admit analytical expressions in terms of plane waves by introducing reflection at the surface through the Fresnel coefficients $\rp$ and $\rs$ for p and s polarization. More precisely, using the identity
\begin{align}
\frac{\ee^{\ii k|\rb-\rb'|}}{|\rb-\rb'|}=\frac{\ii}{2\pi}\int \frac{d^2\Qb}{k_z}\, \ee^{\ii\Qb\cdot(\Rb-\Rb')+\ii k_z|z-z'|},
\label{erridentity}
\end{align}
where $\Rb=(x,y)$ and $k_z=\sqrt{k^2-Q^2+\ii0^+}$ with ${\rm Im}\{k_z\}>0$, operating with $k^2\mathcal{I}_3+\nabla\otimes\nabla$ on the exponentials inside the integrand, and projecting onto the dyadic identity $\mathcal{I}_3=\eh_{\rm p}^\pm\otimes\eh_{\rm p}^\pm+\eh_{\rm s}\otimes\eh_{\rm s}+\kk^\pm\otimes\kk^\pm$ defined in terms of polarization and propagation unit vectors $\eh_{\rm p}^\pm=(\pm k_z\Qb-Q^2\zz)/Qk$, $\eh_{\rm s}=(-Q_y\xx+Q_x\yy)/Q$, and $\kk_{\rm p}^\pm=(\Qb\pm k_z\zz)/k$ with $\Qb=(Q_x,Q_y)$, we find $\mathcal{G}_0(\rb,\rb',\omega)=(\ii k^2/2\pi)\int d^2\Qb\big(1/k_z\big)\ee^{\ii\Qb\cdot(\Rb-\Rb')+\ii k_z|z-z'|}\big(\eh_{\rm p}^\pm\otimes\eh_{\rm p}^\pm+\eh_{\rm s}\otimes\eh_{\rm s}\big)$, where upper (lower) signs must be used for $z>z'$ ($z<z'$); likewise, taking $z,z'>0$ and the surface at $z=0$, the reflection component becomes $\mathcal{G}_{\rm ref}(\rb,\rb',\omega)=(\ii k^2/2\pi)\int d^2\Qb\big(1/k_z\big)\ee^{\ii\Qb\cdot(\Rb-\Rb')+\ii k_z(z+z')} \big(\rp\eh_{\rm p}^+\otimes\eh_{\rm p}^-+\rs\eh_{\rm s}\otimes\eh_{\rm s}\big)$, where downward waves emanating from $z'$ are converted into upward waves reaching $z$ upon reflection at the surface. Introducing these expressions into Eq.\ (\ref{EGp}) and taking the atom to be placed at a distance $z_0$ above the surface (i.e., $\rb'=(0,0,z_0)$), we obtain the field $\Eb(\rb,\omega)=(\ii/2\pi)\int d^2\Qb\big(1/k_z\big) \ee^{\ii\Qb\cdot(\Rb-\Rb')+\ii k_z z}\fb(\Qb)$ for $z>0$, where $\fb(\Qb)=k^2\big[\eh_{\rm p}^+(\ee^{-\ii\varphi_0}\eh_{\rm p}^+\cdot\pb+\ee^{\ii\varphi_0}\rp\eh_{\rm p}^-\cdot\pb)+\eh_{\rm s}(\ee^{-\ii\varphi_0}+\ee^{\ii\varphi_0}\rs)(\eh_{\rm s}\cdot\pb)\big]$ with $\varphi_0=k_zz_0$. In the far-field limit ($kr\gg1$), this expression yields $\Eb(\rb,\omega)\rightarrow\fb(k\Rb/r)\,\ee^{kr}/r$, which allows us to evaluate the emitted power as the integral of the radial component of the Poynting vector over a distant upward hemisphere, $(c/2\pi)\int_{z>0} d\Omega_\rb |\fb(k\Rb/r)|^2$, from which the photon emission rate $\Gamma_{\rm em}$ is obtained by again dividing the result by $\hbar\omega$. Finally, we find
\begin{align}
\Gamma_{\rm em}=&\frac{k^3}{\hbar}\int_0^1 d\mu \bigg[
(1-\mu^2) \left|1+\ee^{2\ii\mu kz_0}\rp\right|^2\,|p_z|^2
\label{Gammaem}\\
&+\frac{1}{2}\left(\left|1+\ee^{2\ii\mu kz_0}\rs\right|^2+\mu^2\left|1-\ee^{2\ii\mu kz_0}\rp\right|^2\right) \nonumber\\
&\;\;\;\;\;\; \times (|p_x|^2+|p_y|^2)
\bigg].
\nonumber
\end{align}
In the absence of a surface ($\rp=\rs=0$), the above expression reassuringly yields $\Gamma_0/2$, indicating that half of the decay rate in free space is accounted for by upward photon emission (and the other half by downward emission). In the presence of the surface, the decay rate takes a more complicated form (obtained for example by integrating the outgoing Poynting vector over a small sphere surrounding the dipole), expressed in terms of the field induced at the position of the dipole as \cite{paper053} $\Gamma_{\rm decay}=\Gamma_0+(2/\hbar){\rm Im}\{\pb^*\cdot\mathcal{G}_{\rm ref}(\rb_0,\rb_0,\omega)\cdot\pb\}$; for our dipole near a planar surface, this leads to the expression
\begin{align}
\Gamma_{\rm decay}=&\Gamma_0+\frac{1}{\hbar}\int_0^\infty QdQ\,
{\rm Re}\bigg\{\frac{\exp(2\ii k_zz_0)}{k_z}
\label{Gammadecay}\\
&\times \left[
2Q^2\rp |p_z|^2+(k^2\rs-k_z^2\rp)(|p_x|^2+|p_y|^2)
\right]\bigg\}.
\nonumber
\end{align}
The $Q>k$ part of this integral involves evanescent waves (i.e., an imaginary normal light wave vector $k_z$) that contribute to the decay through absorption (proportional to ${\rm Im}\{\rp\}$ and ${\rm Im}\{\rs\}$), for example via the emission of plasmons \cite{paper331}; the $Q<k$ part is a combination of absorption and photon emission.

\begin{figure*}
\begin{centering}
\includegraphics[width=1.0\textwidth]{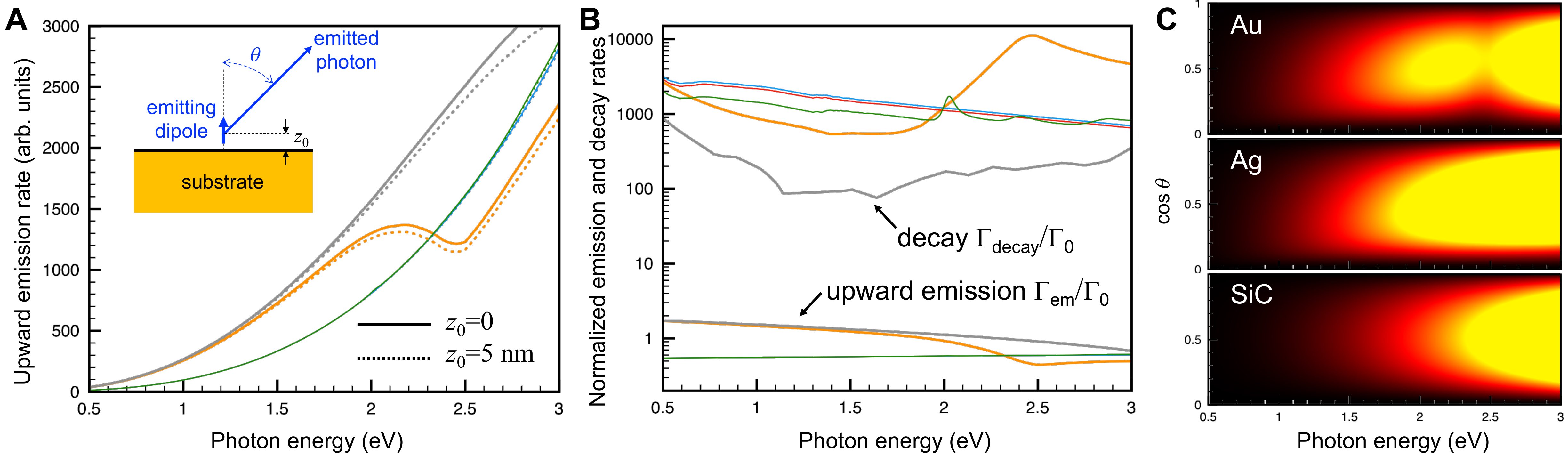}
\par\end{centering}
\caption{{\bf Light emission from an out-of-plane dipole near a planar surface.} ({\bf A}) Spectral dependence of the angle-integrated upward emission rate from a dipole oriented along the normal of Ag, Au, and SiC surfaces. We consider two different dipole-surface distances (see labels). The emission is normalized by the dipole strength $|\pb|^2$. ({\bf B}) Decay and upward emission rates normalized to the emission rate in free space for a dipole separated 2\,nm from the surfaces considered in (A). ({\bf C}) Angle and energy dependence of the upward emission for zero dipole-surface distance. We further consider SiC covered with a monolayer of graphene or with graphene an a monolayer of WS$_2$, which produce very similar results compared with the bare SiC surface (nearly indistinguishable curves).}
\label{planar-surface}
\end{figure*}

\subsection{Emission and decay from a dipole near a planar surface}

We first illustrate the application of Eq.\ (\ref{Gammaem}) to study the emission from a dipole on a planar surface in the absence of a tip. We consider an out-of-plane dipole, which will later be identified with the tip-sample current, as we argue above. Figure\ \ref{planar-surface}A shows the resulting spectra for different materials. A first observation is that the emission grows with photon energy, as expected from the $k^3$ coefficient in front of the integral in Eq.\ (\ref{Gammaem}); in physical terms, the dipole appears to be bigger in front of the photon wavelength as the energy increases, therefore undergoing better coupling to radiation. A second observation is that the emission is similar in magnitude in all cases, and in particular, the results for SiC are nearly indistinguishable when the material supports monolayers of graphene and WS$_2$. A third observation is that the emission rate is almost unchanged when the dipole is separated by 5\,nm from the surface (note that the results are normalized to the dipole strength $|\pb|^2$), as this distance is much smaller than the light wavelength in the spectral range under consideration.\\

The emission rate is only a part of the decay rate, as the latter also receives contributions from absorption by the material [see Eq.\ (\ref{Gammadecay})]. At zero separation, the decay rate diverges as a result of the unphysical $1/r$ Coulomb interaction at small distances in the local response approximation used here to describe the materials (i.e., we use frequency-dependence dielectric functions). This divergence is however removed when incorporating spatial dispersion in the response, which is in general an effect that becomes important only at small separations below 1\,nm \cite{paper119}, so, for simplicity, we ignore it in this discussion. In Fig.\ \ref{planar-surface}B, we plot the upward emission and decay rates when the dipole is 2\,nm above the surface, both of them normalized to the emission rate in vacuum $\Gamma_0$. The normalized emission rate near the material takes unity-order values, and therefore, we conclude that the presence of the surface does not significantly affect the emission relative to free space, as we had already anticipated by the close resemblance of the results obtained for different materials in Fig.\ \ref{planar-surface}A. At low frequencies, gold and silver produce good screening (perfect-conductor limit), thus doubling the magnitude of the dipole (through its image contribution) and increasing the upward emission by a factor of 4 relative to free space. In contrast, the decay rate becomes 2-4 orders of magnitude larger than in free space, with silver giving the lowest values among the materials under consideration because of the relatively low losses in this noble metal. Again, the decay rate in SiC does not change significantly (in log scale) when decorating it with monolayers of graphene or WS$_2$. The decay is thus dominated by non-radiative contributions [parallel wave vector $Q>k$ in Eq.\ (\ref{Gammadecay})]. Although this is an inelastic contribution, it should result in a dark tunneling current, which is still smaller than the elastic tunneling rate (see below). The radiationless absorption accompanying this process should additionally produce local heating, which raises the interesting question of whether it can be detected through its bolometric effect, or perhaps \textit{via} direct charge-carrier separation.\\

The angular dependence of the emission (Fig.\ \ref{planar-surface}C) does not depart significantly from the $\cos^2\theta$ distribution in free space as a function of emission angle relative to the dipole orientation. This dependence should also change when a tip is present, although the bulk of the emission is directed sideways, and therefore, the effect of the tip should not be dramatic.

\begin{figure}
\begin{centering}
\includegraphics[width=0.4\textwidth]{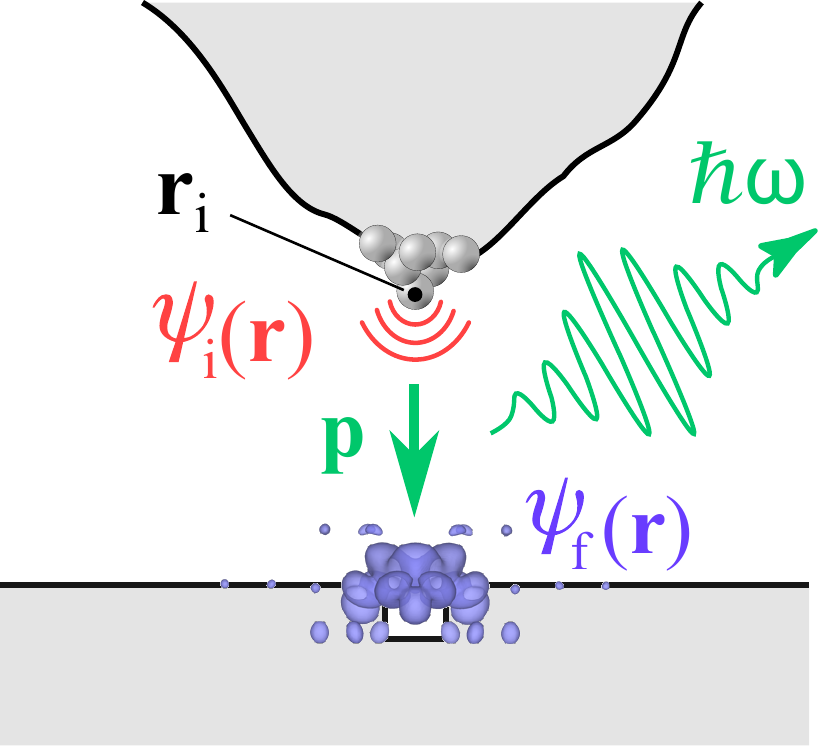}
\par\end{centering}
\caption{{\bf Schematic representation of the elements involved in the theoretical description of STM-induced light emission from a defect in a 2D material.} A dipole moment (downward arrow) associated with the transition between initial and final electron states acts as an electromagnetic source and producing light emission away from the tip region, assisted by coupling to tip plasmons. The initial state originates in an atomic protuberance from the tip and therefore, it can be approximated as an spherical wave emanating from a tip position $\rb_{\rm i}$.}
\label{sketch}
\end{figure}

\subsection{Photon emission induced by tunneling into a defect in a 2D material}

We now address the question of what is actually measured by recording maps of light emission in a STM while scanning a defect in a 2D material. We adopt the dipole approximation [Eqs.\ (\ref{EGp}) and (\ref{pb})] and consider a spherical evanescent electron wave $\psi_{\rm i}(\rb)=C\ee^{-\kappa_{\rm i}|\rb-\rb_{\rm i}|}/|\rb-\rb_{\rm i}|$ emanating from the tip \cite{TH1985} (centered at $\rb_{\rm i}$, see Fig.\ \ref{sketch}), although the conclusions drawn below do not depend on the exact details of the initial wave function, but rather on its atomic-scale origin. Here, $C$ is a normalization constant that depends on the detailed atomic shape and composition of the tip, while $\kappa_{\rm i}$ gives the evanescent spill out of the initial state outside the tip, which is determined by its binding energy relative the the vacuum threshold (see below). From the representation of this type of wave given by Eq.\ (\ref{erridentity}) combined with Eq.\ (\ref{pb}), we find a transition dipole
\begin{align}
\pb=\frac{-e\hbar C}{2\pi\me\omega}\int d^3\rb\,\psi^*_{\rm f}(\rb)
\int d^2\kparb\; &\ee^{\ii\kparb\cdot(\Rb-\Rb_{\rm i})-\kappa_z(z_{\rm i}-z)}
\nonumber\\ &\times\left(\ii\kparb/\kappa_z+\zz\right),
\nonumber
\end{align}
where $\kappa_z=\sqrt{\kappa_i^2+\kpar^2}$, we have renamed the integration variable as $\Qb\rightarrow\kparb$ to distinguished the optical parallel wave vector $\Qb$ (see above) from the electronic parallel wave vector $\kparb$, and we have used the fact that $z<z_{\rm i}$ in the region near the final state (see Fig.\ \ref{sketch}). We now argue that the initial electron evanescent wave has a decay length $1/\kappa_{\rm i}\lesssim\hbar/\sqrt{2\me\phi}$ dictated by the tip work function $\phi\sim5\,$eV; this leads to $1/\kappa_{\rm i}\lesssim0.1\,$nm, which we have to compare with the lateral size of the 2D final state $D\sim1\,$nm; we conclude that $\kappa_{\rm i}D\ll1$, and therefore, the largest values of $\kpar\sim1/D$ needed in the above integral to obtain a good representation of the final state can be neglected in front of $\kappa_{\rm i}$; we can thus approximate $\kappa_z\approx\kappa_{\rm i}$ and disregard the in-plane components of $\pb$, which then reduces to
\begin{align}
\pb\approx\frac{-2\pi e\hbar C}{\me\omega}\int dz\,\psi^*_{\rm f}(\Rb_{\rm i},z)\, \ee^{-\kappa_{\rm i}(z_{\rm i}-z)}\,\zz.
\label{nearlyfinal1}
\end{align}
As the emission is proportional to $|\pb|^2$ (see Eq.\ (\ref{Gammaem})), we conclude that the photon yield is probing the final-state wave function at the lateral position of the atomic-scale tip: $\Gamma_{\rm em}\propto\big|\int dz\,\psi_{\rm f}(\Rb_{\rm i},z)\,\ee^{-\kappa_{\rm i}(z_{\rm i}-z)}\big|^2$.\\

We can obtain a more insightful result by noticing that the $\kparb$ in-plane Fourier component of the final state must decay with distance $z$ above a plane at $z=z_f$ that is away from the sample as $\exp\left[-\sqrt{\kappa_f^2+\kpar^2}(z-z_f)\right]\approx\exp(-\kappa_f(z-z_f))$, where $\kappa_f$ is determined by the final state energy relative to the vacuum level, and in the rightmost part of this expression we have approximated $\kpar\ll\kappa_f$, similar to what we have done for $\kappa_i$ in the initial state. Also, neglecting the photon energy and applied bias potential energy relative to the binding energy of initial and final states referred to vacuum, we further approximate $\kappa_f\approx\kappa_i$, which allows us to work the integral in Eq.\ (\ref{nearlyfinal1}) to find
\begin{align}
|\pb|^2=\left(\frac{2\pi e\hbar C}{\me\omega}\right)^2\,d\,\ee^{-2\kappa_id}|\psi_f(\Rb_i,z_f)|^2,
\nonumber
\end{align}
where $d=z_i-z_f$ is the tip-sample distance. The emission probability is then obtained as $4k^3|\pb|^2/3\hbar$ (see expression for $\Gamma_0$ above) multiplied by the radiative Purcell factor (i.e., the ratio of the radiative component of the local density of optical states to its value in vacuum). This factor can be substantially enhanced due to coupling to plasmons.
\\

We now recall that the elastic tunneling current is also proportional to $\big|\psi_{\rm f}(\Rb_i,z_f)\big|^2$ \cite{TH1985}, and therefore, both the elastic current and the inelastic photon emission rate are proportional to the final state electron probability under the tip position. In more detail, specifying the Tersoff and Hamann \cite{TH1985} formalism to a final state with the characteristics considered above, we find the STM elastic current to be contributed by the initial state under consideration with the matrix element
\begin{align}
|M|^2=\left(\frac{2\pi\hbar^2 C}{\me}\right)^2\,\ee^{-2\kappa_id}|\psi_f(\Rb_i,z_f)|^2.
\nonumber
\end{align}
We thus conclude that both STM and STML intensities are proportional to the defect orbital electron probability $|\psi_f(\Rb_i,z_f)|^2$ at the sample surface, and both of them are attenuated by the same exponential factor $\ee^{-2\kappa_id}$.


\begin{figure*}
\begin{centering}
\includegraphics[width=0.7\textwidth]{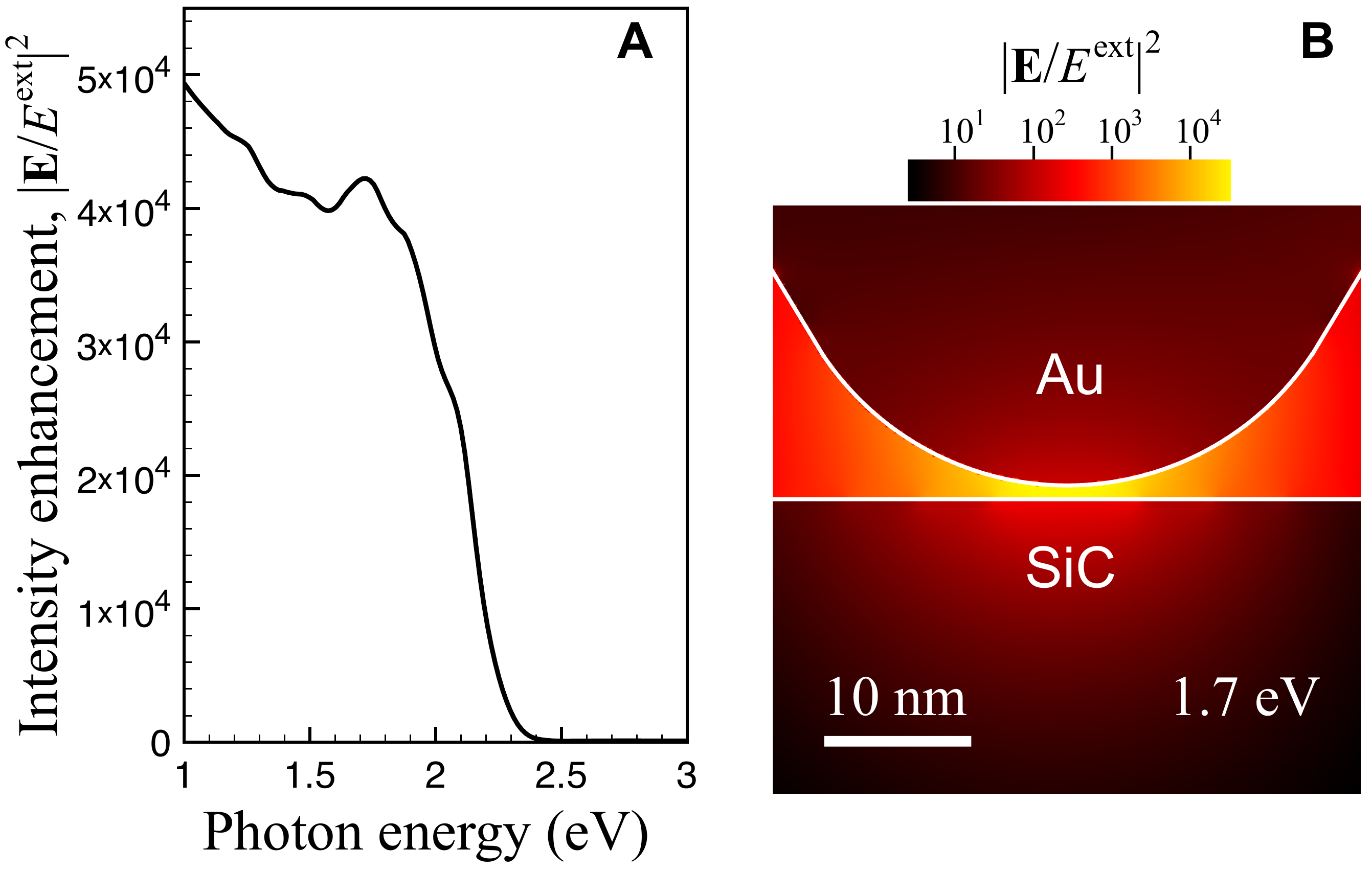}
\par\end{centering}
\caption{{\bf Near-field enhancement at the tip.} We illustrate this effect by considering a Au conical tip (20\,nm radius, $30^\circ$ semi-angle) near a SiC surface (1\,nm gap) irradiated with p-polarized light (external field amplitude $E^{\rm ext}$) under $45^\circ$ incidence relative to the tip axis. ({\bf A}) Spectral dependence of the intensity enhancement at the center of the tip-sample gap. ({\bf B}) Near-field intensity distribution for 1.7 eV light in a plane containing the tip axis (nearly independent on the azimuthal orientation of the incidence direction). Results are averaged over the vertical tip extension from 1 to $2\,\mu$m.}
\label{nearfield}
\end{figure*}

\subsection{Field enhancement produced by the tip}

The inelastic tunneling dipole $\pb$ calculated in the previous section acts as the source of radiation emission observed in STM luminescence. The calculations presented above for a dipole emitting near a planar surface already provide a crude approximation of some of these characteristics of the resulting emission rates, which were estimated to be similar to those of the dipole in free space. However, the emitting dipole in STML sits at an electromagnetic {\it hotspot} produced by the relatively sharp morphology of the tip -- a geometry that generally gives rise to large near-field enhancements in metals, mediated by plasmons that propagate on the surface of the material.\\

An intuitive understanding of the enhancement produced in the emission by the presence of the tip is provided by the reciprocity theorem \cite{J99}, which states that the component of electric field generated at a given position $\rb$ along a direction $\hat{\ub}$ by a dipole sitting at $\rb'$ and oriented along $\hat{\ub}'$ must coincide with the component of the field produced at $\rb'$ along the direction $\hat{\ub}'$ by a dipole sitting at $\rb$ and oriented along $\hat{\ub}$ in the absence of magnetic and nonlinear optical effects (i.e., for the standard linear response in the materials here considered). In more rigorous terms, this is a consequence of the symmetry of the electromagnetic Green tensor, which under those conditions satisfies $\mathcal{G}(\rb,\rb',\omega)={\rm transpose}\{\mathcal{G}(\rb',\rb,\omega)\}$.\\

A direct application of the reciprocity theorem to our STML geometry (i.e., with $\rb'$ at the tip and $\rb$ at the light detector) allows us to state that the enhancement in the emission rate from the transition dipole of Fig.\ \ref{sketch} along a given outgoing direction must be equal to the enhancement of the near-electric-field intensity at the position of that dipole for light incident from that same direction. The enhancement value is the so-called Purcell factor $P(\omega)$ \cite{P1946}, which also determines the emission rate from an optical emitter (e.g., a quantum dot of a fluorescent molecule). We calculated $P(\omega)=|\Eb/E^{\rm ext}|^2$ (i.e., the ratio of local to externally incident field intensities, which must be then understood as the emission enhancement through reciprocity) in Fig.\ \ref{nearfield} for a tip of 20\,nm radius and 1\,nm separation from a SiC surface. The enhancement reaches 4 orders of magnitude at energies below the gold plasmon (i.e., $<2.5\,$eV, see Fig.\ \ref{nearfield}A) and is highly localized near the tip (Fig.\ \ref{nearfield}B).\\

The spectral profile of the enhancement strongly depends on the detailed tip morphology. Upon examination of several tips (see below), it is unlikely that a spectrally narrow plasmon is supported by the tip. However, similar to the one considered in the calculations of Fig.\ \ref{nearfield}, the tip is expected to act as a collector of light that couples to propagating plasmons, which in turn move toward to tip region, thus enhancing the field relative to the incident one. In this respect, there is room for improvement of tip design in STML, for example by decorating it with in-/out-coupling elements, such as gratings carved far from the tip region. Additionally, in W tips coated with Au, such as those used in this study, finite penetration of the optical field beyond the Au skin depth ($\sim20\,$nm) can result in an increase of absorption in the low-energy region, where W is particularly lossy. This effect could explain why our observed emission profiles are peaked in the $1-2$\,eV region, with Au plasmons acting only below $\sim2.5$\,eV and W absorption taking place below $\sim1\,$eV.

\subsection{Relative magnitude of elastic and inelastic tunneling currents}

A rough estimate of the ratio of photons emitted per electron tunneled to a given final state $\psi_\text{f}$ is readily obtained by dividing $\Delta\omega\times(4k^3|\pb|^2/3\hbar)\,P(\omega)$ by $(2\pi/\hbar)|M|^2$, where $\Delta\omega$ is the emission bandwidth. Notice that we have introduced the Purcell factor $P(\omega)$ in the emission rate to include the near-field enhancement effect at the tip, as we discuss above. With the above approximations, the result is independent of tip position and the ratio reduces to $(3\pi\alpha/2)(d/\lambda)^2(\Delta\omega/\omega)$, where $\alpha\approx1/137$ is the fine-structure constant and $\lambda$ is the emission wavelength; considering photon emission centered around $\hbar\omega\sim2\,$eV with bandwidth $\Delta\omega\sim1$\,eV, a tip-sample distance $d\sim1\,$nm, and a Purcell factor $P\sim10^4$ (see Fig.\ \ref{nearfield}), we find a photon-to-electron overall ratio of $10^{-4}$, in excellent agreement with our experimentally estimated yield. Incidentally, we are neglecting in this calculation inelastic tunneling associated with nonradiative processes (i.e., mediated by direct material absorption), which could produce a significant contribution (see Fig.\ \ref{planar-surface}B), although this should not change the order of magnitude of the estimated ratio given above.\\

It is important to note that the elastic tunneling requires the energy $E_{\rm 2D}$ of the final 2D defect state studied in this work to be below the Fermi level of the metallic tip $E_{\rm F}^{\rm tip}$, while in STM luminescence the photon energy $\hbar\omega$ compensates for the difference between tip and 2D states, which must then satisfy the condition $0<\hbar\omega<E_{\rm F}^{\rm tip}-E_{\rm 2D}$, thus producing a correspondingly broad spectral emission. Obviously, the difference $E_{\rm F}^{\rm tip}-E_{\rm 2D}$ depends on the applied potential energy $V_{\rm bias}$, with the onset for emission determined by the condition $V_{\rm bias}>E_{\rm 2D}-E_{\rm F}^{\rm 2D}$ (i.e., the tip Fermi energy must be above the 2D defect state energy).

\section{Monolayer and bilayer \ce{WS2} on Gr/SiC sample}
Mono- and bilayer \ce{WS2} islands were grown \textit{ex-situ} by chemical vapor deposition~\cite{Kastl2017cvd} on epitaxial graphene on (6H)-SiC substrates~\cite{emtsev2009towards}. Further details can be found in Refs.~\citenum{Kastl2017cvd,schuler2019overview}. We identified several types of defects in as-grown samples including transition metal substitutions and oxygen substituting for sulfur~\cite{schuler2019overview,barja2018identifying}. Sulfur vacancies, which are absent in as-grown samples can be deliberately introduced by annealing the sample at \unit{600}{$^\circ$C} in vacuum~\cite{schuler2018large}.\\

\section{STM/STS and CO-tip nc-AFM measurements}
All scanning probe measurements were conducted at low-temperature ($T \approx 6\,\text{K}$) and ultra-high vacuum ($p\approx 10^{-10}\,\text{mbar}$).
Scanning tunneling microscopy (STM) images were recorded using constant-current feedback. Scanning tunneling spectroscopy (STS) spectra were performed in constant-height mode with a \unit{5}{mV} lock-in amplitude. The bias was applied to the sample.\\

For the noncontact atomic force microscopy (nc-AFM) measurements a qPlus~\cite{Giessibl1999} quartz-crystal cantilever was employed (resonance frequency $f_{0}\approx 30\,\text{kHz}$, spring constant $k\approx 1800\,\text{N/m}$, quality factor $Q\approx 30,000$, and oscillation amplitude $A\approx 1\,\text{\AA}$). The metallic tip was functionalized with a CO molecule for enhanced resolution~\cite{Gross2009a,Mohn2013}.\\

\begin{figure*}[]
\includegraphics[width=\textwidth]{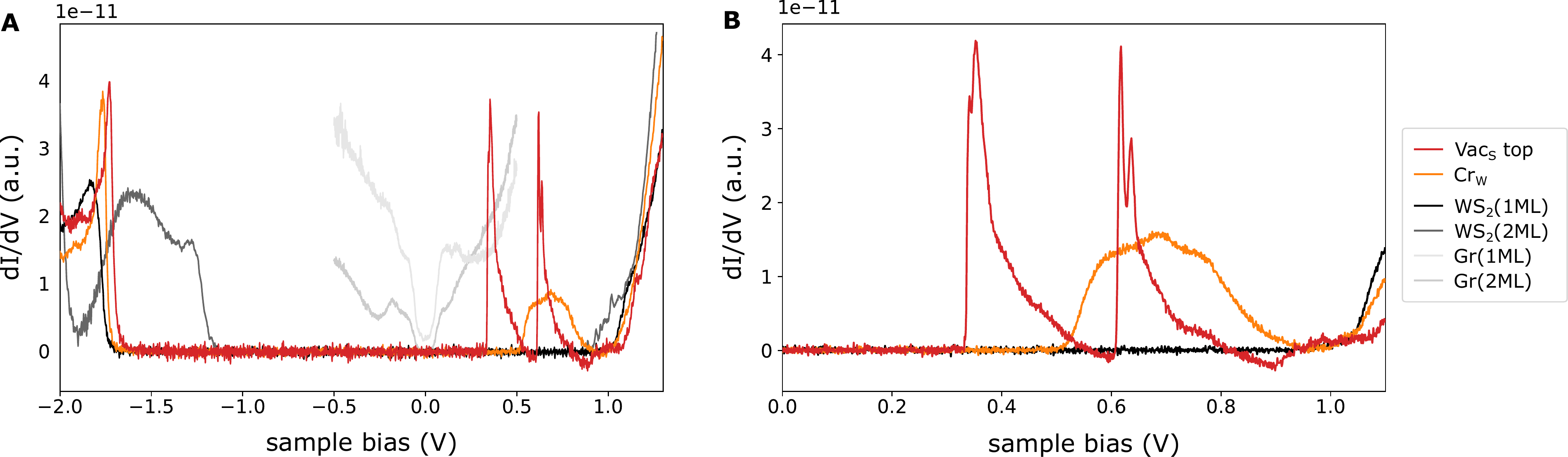}
\caption{\label{fig:dIdV_long}
\textbf{dI/dV spectroscopy of \ce{WS2} defects and substrates.} \textbf{A} dI/dV spectroscopy on different substrates and defect locations (see legend on the right). \textbf{B} dI/dV spectroscopy of the two unoccupied Vac$_\text{S}$ top (red), and three Cr$_\text{W}$ defect states (orange). 
}
\end{figure*}

In \figref{fig:dIdV_long}, dI/dV spectra at different sample locations are shown: Mono- and bilayer epitaxial graphene on SiC (light gray), mono- and bilayer \ce{WS2} on Gr(2ML)/SiC (black and dark gray), a sulfur vacancy (Vac$_\text{S}$) and chromium substituent (Cr$_\text{W}$) in \ce{WS2}(1ML)/Gr(2ML)/SiC. Vac$_\text{S}$ features two unoccupied defect states in the band gap~\cite{schuler2018large} and Cr$_\text{W}$ three defect states close to the conduction band minimum~\cite{schuler2019overview}.

\section{STM luminescence}
For the STML measurements we used constant-current STM feedback to drive a constant electron flux at varying tunneling biases at a specific location on the sample. The emitted light was collimated using a high numerical aperture, achromatic lens close to the tunneling junction. Outside the vacuum chamber the light was refocused into an optical fiber to guide it to a point detector or spectrometer. A photon multiplier tube (Hamamatsu, H7421-50) and a grating spectrometer (Princeton Instruments, SpectraPro 2300i) with a thermoelectrically-cooled charge coupled device photon detector (Andor, Newton) were used. The STML spetra were measured using a 150 grooves per millimeter grating and were not corrected for the wavelength dependent detection efficiency.

\subsection{Optical spectra at different sites}

\begin{figure*}[]
\includegraphics[width=0.9\textwidth]{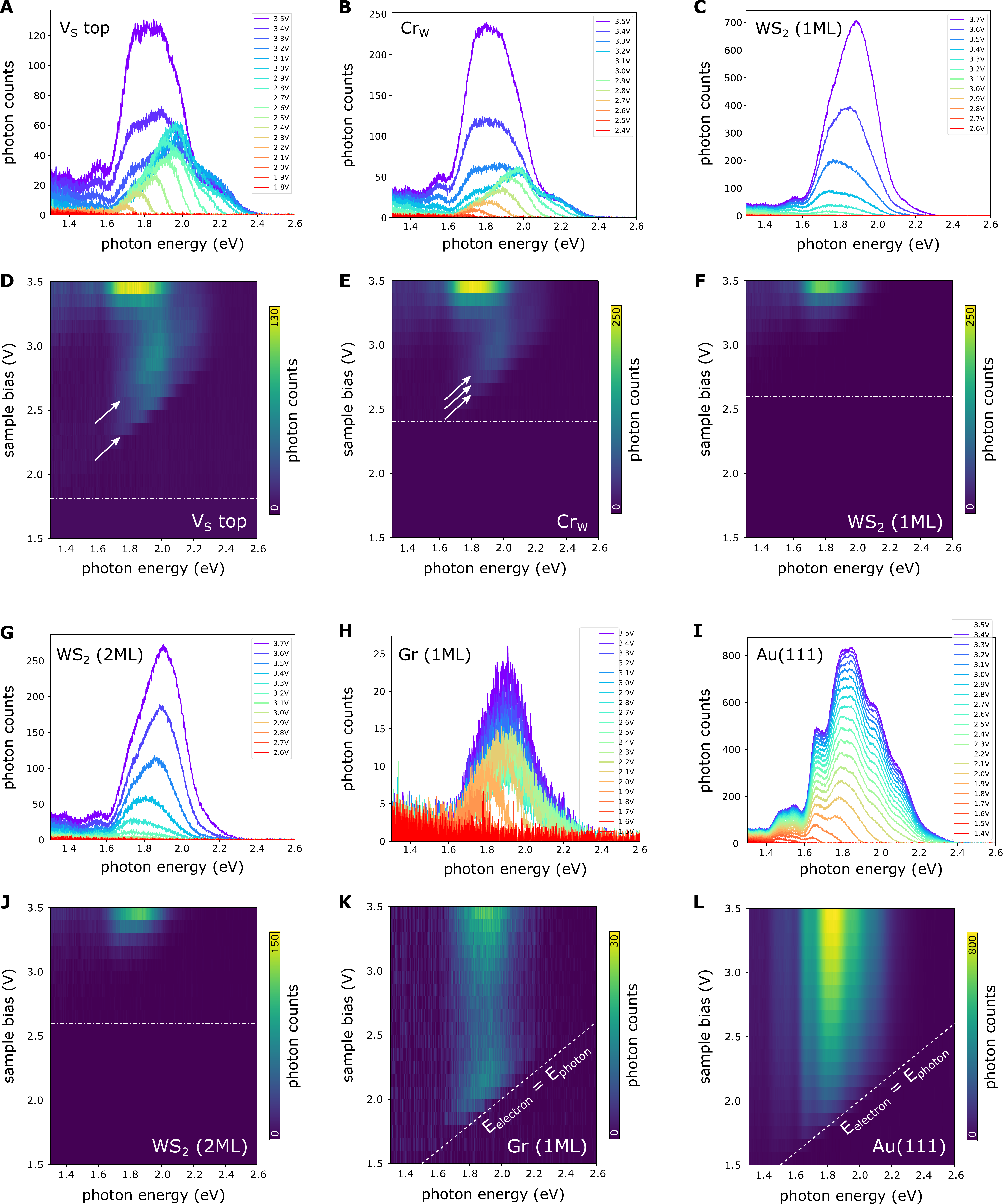}
\caption{\label{fig:OpticalSpectraAll}
\textbf{Spectrally-resolved STML emission at different sites.} \textbf{A-L} STML emission spectrum at different tunneling biases recorded on a Vac$_\text{S}$ top (\textbf{A},\textbf{D}), a Cr$_\text{W}$ (\textbf{B},\textbf{E}), 1ML \ce{WS2} (\textbf{C},\textbf{F}), 2ML \ce{WS2} (\textbf{G},\textbf{J}), 1ML graphene on SiC (\textbf{H},\textbf{K}) and Au(111) (\textbf{I},\textbf{L}) using the same tip. 
The arrows in \textbf{D} and \textbf{E} indicate the calculated maximum photon energy from transitions involving the unoccupied defect states shown in \figref{fig:dIdV_long}B as final states. 
The inclined dashed lines in \textbf{K} and \textbf{L} indicate the isoenergetic line where the photon energy and the electron energy are equal.
At sample biases smaller than indicated by the horizontal dash-dotted no spectra were recorded because no emission was observed.
}
\end{figure*}

The STM luminescence is specific to which atomic site electrons are injected. In \figref{fig:OpticalSpectraAll} the STML spectra recorded on Vac$_\text{S}$ top, Cr$_\text{W}$, non-defective \ce{WS2}(1ML), \ce{WS2}(2ML), Gr(1ML)/SiC and Au(111) are shown. The spectra positions (except for the Au surface) were located within a few \unit{100}{nm} away from each other and were measured with the exact same tip (including the Au). The two defects Vac$_\text{S}$ top and Cr$_\text{W}$ introduce an additional emission band at tunneling biases below the \ce{WS2} bulk emission. The steps in the defect emission (white arrows) can be explained by the discrete defect states, which are the final states of the inelastic electron tunneling process. On both \ce{WS2}(1ML) and \ce{WS2}(2ML) only emission at higher tunneling biases is observed (top left corner in \figref{fig:OpticalSpectraAll}F,J), which correspond to inelastic electron tunneling into the \ce{WS2} conduction band. Also for Gr/SiC and Au(111) STML is observed. The detected photons have energies lower or equal to the injected electron energy. We associate the isoelectronic transitions where the photon energy equals the electron energy (inclined dashed lines in \figref{fig:OpticalSpectraAll}K,L) to inelastic transitions to unoccupied states at the Fermi level.\\

\subsection{STML photon maps of Vac$_\text{S}$ and Cr$_\text{W}$}

\begin{figure*}[]
\includegraphics[width=\textwidth]{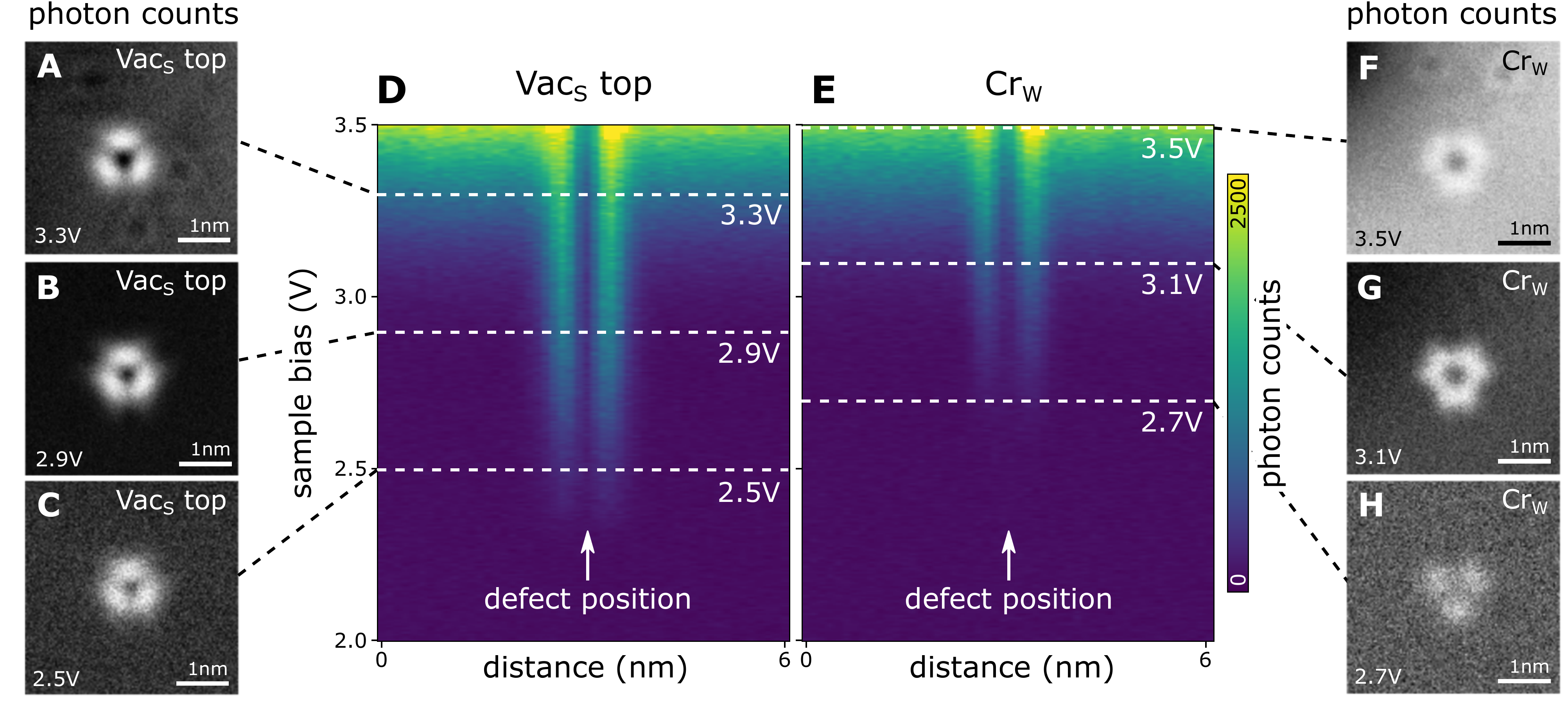}
\caption{\label{fig:VacancyVsCrW}
\textbf{Vac$_\text{S}$ top and Cr$_\text{W}$ photon maps.} \textbf{A-C} Spectrally integrated photon maps of Vac$_\text{S}$ top at \unit{3.3}{V} (A), \unit{2.9}{V} (B) and \unit{2.5}{V} (C). \textbf{D} Spectrally integrated photon emission across Vac$_\text{S}$ top as a function of tunneling bias. \textbf{D} Spectrally integrated photon emission across Cr$_\text{W}$ as a function of tunneling bias. \textbf{F-H} Photon maps of Cr$_\text{W}$ at \unit{3.5}{V} (F), \unit{3.1}{V} (G) and \unit{2.7}{V} (H). The photon maps of Vac$_\text{S}$ top and Cr$_\text{W}$ resemble their respective defect orbitals.
}
\end{figure*}

In \figref{fig:VacancyVsCrW} the spectrally integrated photon maps of Vac$_\text{S}$ top and Cr$_\text{W}$ at different biases are shown. The shape and spatial extent of the STML map resemble their respective in-gap defect orbitals.\\

\subsection{STML at negative sample bias}

\begin{figure*}[]
\includegraphics[width=0.8\textwidth]{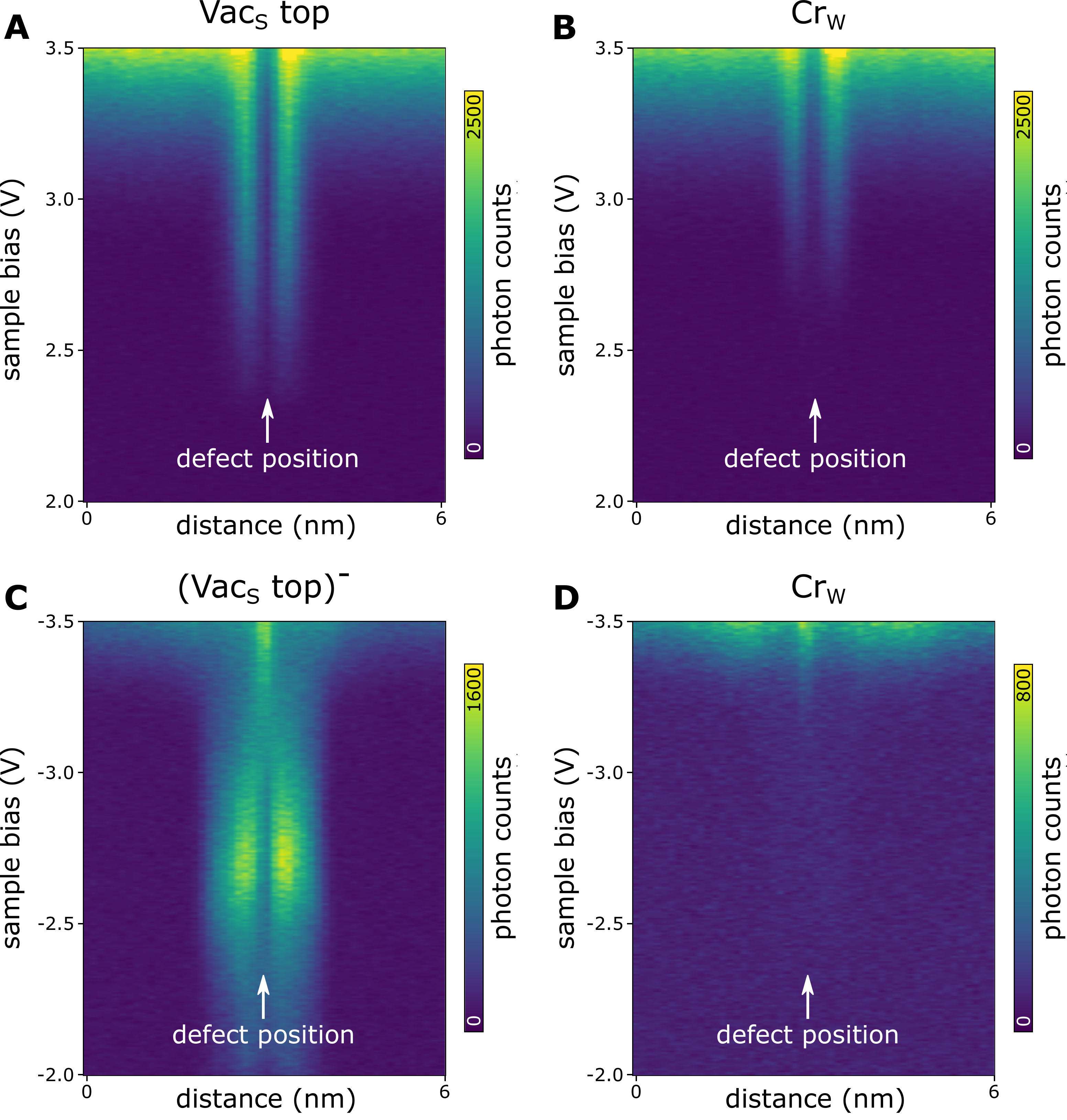}
\caption{\label{fig:EmissionVsBiasNeg}
\textbf{Photon emission vs position at negative bias.} \textbf{A-D} Spectrally integrated photon maps of a Vac$_\text{S}$ top (\textbf{A},\textbf{C}) and a Cr$_\text{W}$ defect (\textbf{B},\textbf{D}) at positive (\textbf{A},\textbf{B}) and negative (\textbf{C},\textbf{D}) sample bias. STML is observed for both bias polarities. Note that the sulfur vacancy gets negatively charged at high negative sample bias because tip-induced band bending populates the formerly unoccupied defect state close to the Fermi energy.
}
\end{figure*}

STML is also observed at negative sample bias. In \figref{fig:EmissionVsBiasNeg} the spectrally integrated photon emission across Vac$_\text{S}$ top and Cr$_\text{W}$ for positive and negative tunneling biases is compared. The bias onset of STML emission depends on the tunneling polarity. At negative biases electrons from \textit{occupied} states in the sample tunnel inelastically to unoccupied tip states. Cr$_\text{W}$ does not feature any occupied defect states in the band gap, hence no additional defect emission band is observed at negative bias (\figref{fig:EmissionVsBiasNeg}D). For Vac$_\text{S}$ the case is not so simple because tip-induced band bending pulls the lowest unoccupied defect state below the substrate Fermi level at biases exceeding \unit{-1.2}{V}~\cite{schuler2018large}. Therefore, the defect becomes negatively charged (Vac$_\text{S}^-$) in the vicinity of the tip. Accordingly, STML for Vac$_\text{S}$ at negative bias (\figref{fig:EmissionVsBiasNeg}C) is likely related to inelastic tunneling events out of the populated (formerly unoccupied) defect states.

\subsection{STML current dependence}

\begin{figure*}[]
\includegraphics[width=0.9\textwidth]{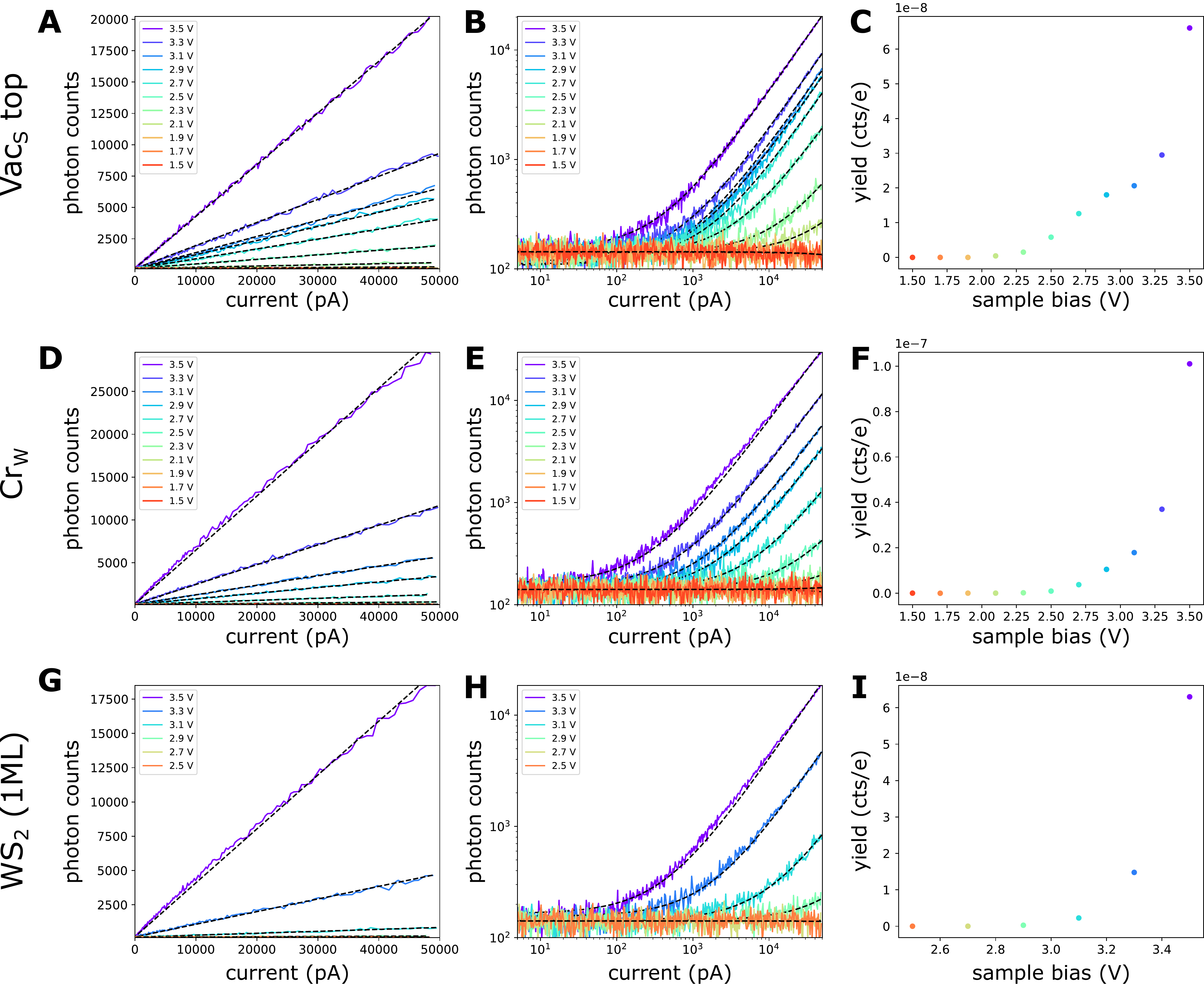}
\caption{\label{fig:EmissionVsCurrent}
\textbf{Photon counts vs tunneling current.} \textbf{A-I} The first two columns display the photon counts as a function of tunneling current for different biases. The linear relation (linear fits indicated by dashed black lines) between photon counts and injected electrons suggests a single electron process. 
In the last column the extrinsic emission yield obtained from the slope of the linear fit is plotted as a function of sample bias. This relation resembles the counts vs bias traces (Fig.~2F in the main manuscript) since they were recorded at constant current. The intrinsic quantum yield of the inelastic tunneling process can be obtained by accounting for setup related losses. 
}
\end{figure*}

The photon counts follow a linear relation as a function of tunneling current at least up to \unit{50}{nA}. The linear dependence between injected electrons and emitted photons suggest a single-electron excitation process. 
No saturation behavior or current-dependent change in emission spectrum was observed like reported for other systems~\cite{merino2018bimodal}.
The extrinsic emission yield (detected photons per tunneling electron, $\eta_\text{exp}$) is a product of the intrinsic quantum efficiency $\eta_\text{0}$ of the radiative tunneling process, the tip-mediated coupling efficiency $\kappa_\text{tip}$ from the tunnel junction into the far-field (plasmon enhancement) and the detection efficiency of the optical setup $\kappa_\text{setup}$. 

\begin{equation} \label{eq:yield}
 \eta_\text{exp} = \eta_\text{0}\kappa_\text{tip}\kappa_\text{setup}
\end{equation}

The setup detection efficiency is about $10^{-3}$, which accounts for the solid angle of collection, optical losses (fiber coupling, mirrors, lenses) and the detector quantum efficiency. At larger tunneling biases of about \unit{3.5}{V}, on the order of 10$^{-7}$ photons per electron are detected in the far field. Hence the intrinsic quantum efficiency times the tip enhancement is estimated as $10^{-4}$ using a standard Au coated W tip. As discussed in the next section, the tip shape has a decisive impact on the brightness and spectral shape of the emission. Here we rely on stochastic tip changes by nanometer deep indentations into a Au surface. The plasmonic coupling of the quantum emitter might be easily enhanced by choosing a more optimized plasmonic or optical cavity.

\section{Tip preparation}~\label{sec:tip}
For the STM/STS and STM luminescence measurements we used an etched tungsten tip wire. Inside the STM the tip was sharpened by field emission at high biases (\unit{100}{V}) and nanometer mechanical pokes into a Au(111) surface. This procedure serves two purposes: (i) to form an atomically-sharp tip apex for tunneling and (ii) a mesoscopic Au coating of the tip for plasmonic field enhancement.\\

\begin{figure*}[]
\includegraphics[width=0.6\textwidth]{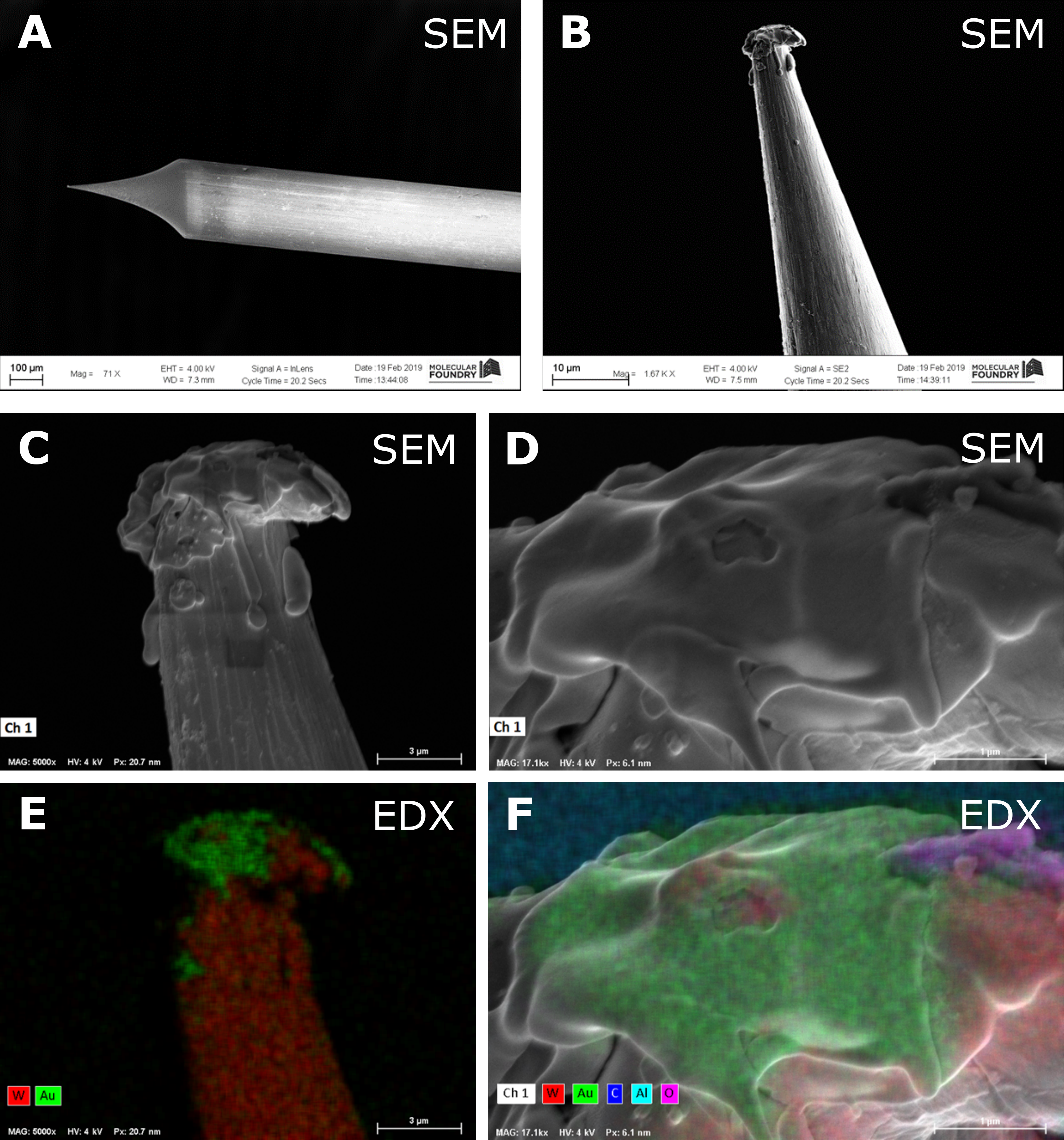}
\caption{\label{fig:SEM_EDX}
\textbf{STM tip analysis.} \textbf{A,B} SEM images of a representative etched W tip after field emission and indentations into a Au surface. \textbf{C,D} SEM close-up on the tip apex. \textbf{E,F} Corresponding EDX elemental analysis at the same position. The tip apex is coated with Au.
}
\end{figure*}

A representative STM tip was analyzed using scanning electron microscopy (SEM) and energy-dispersive X-ray spectroscopy (EDX). We used the ZEISS Ultra 55 FESEM equipped with an Bruker X-ray energy dispersive spectrometer for elemental mapping. As seen in \figref{fig:SEM_EDX} the tip apex becomes morphologically less defined on the sub-micron scale after field emission and nano-indentations into the Au surface. However, the W tip wire is clearly coated with a Au film at the very apex of the tip, which results in a plasmonic enhancement effect.\\
 
\begin{figure*}[]
\includegraphics[width=\textwidth]{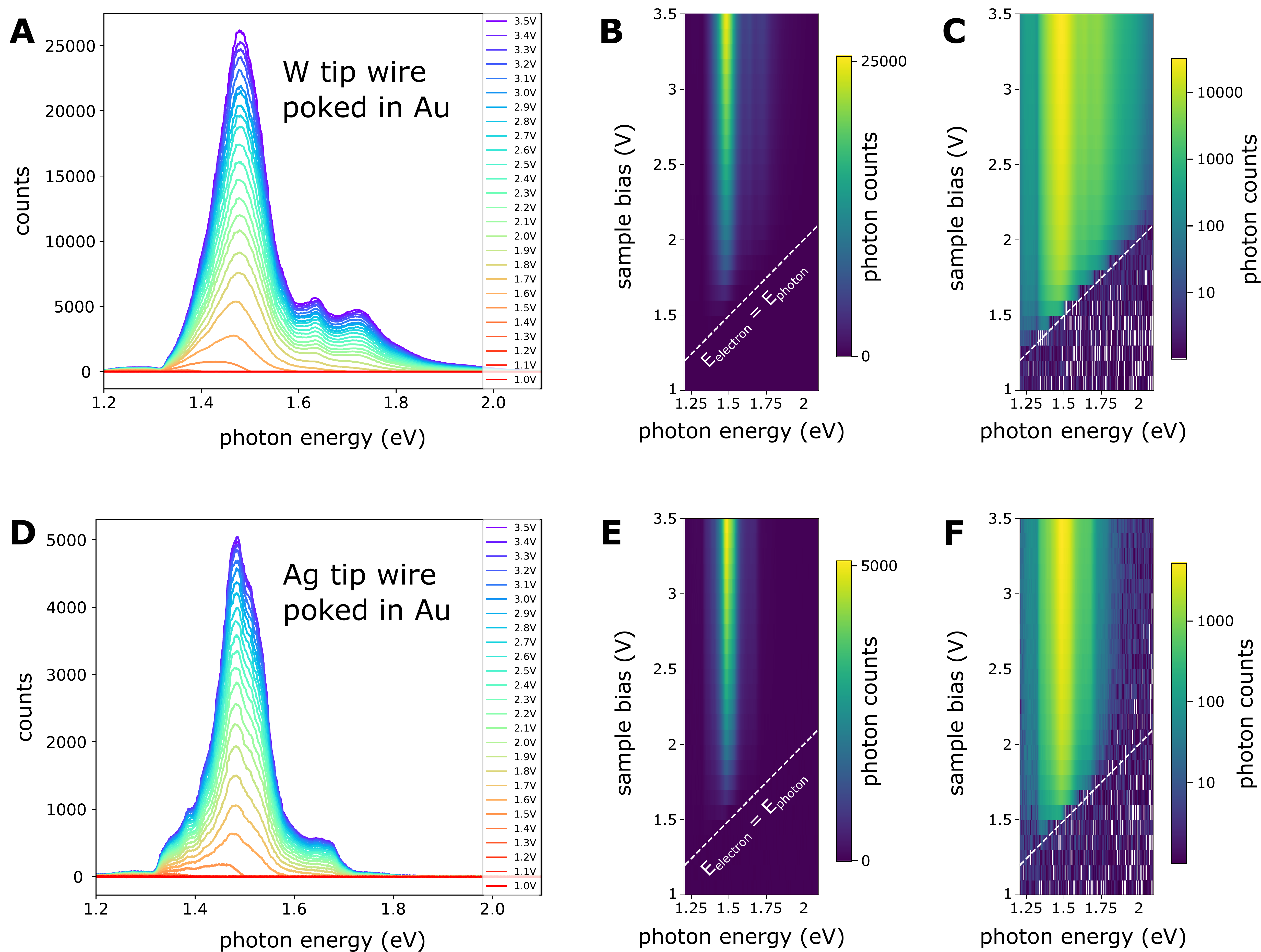}
\caption{\label{fig:WvsAg}
\textbf{Tungsten vs silver tip wire.} \textbf{A-C} STML spectra recorded on a Au(111) surface using an etched W wire after tip treatment on Au. \textbf{D-F} STML spectra recorded on a Au(111) surface using an etched Ag wire after tip treatment on Au. The emission spectrum is very similar for both the W and Ag tip after tip shaping on Au. Hence, the spectral modulation of the STML emission is mainly dominated by the tip shape and material of the mesoscopic tip apex.
}
\end{figure*}

We also compared etched tungsten tips to etched silver tips. Like for the W tips, we used field emission and surface pokes in Au to sharpen the Ag tip. The bulk tip materials has essentially no effect on the STM luminescence spectrum as seen in \figref{fig:WvsAg}. The STML spectra on Au(111) are very similar for both the W and Ag tips. This suggests that only the mesoscopic tip shape and material (Au in both cases) matters for the spectral emission properties.\\

\begin{figure*}[]
\includegraphics[width=\textwidth]{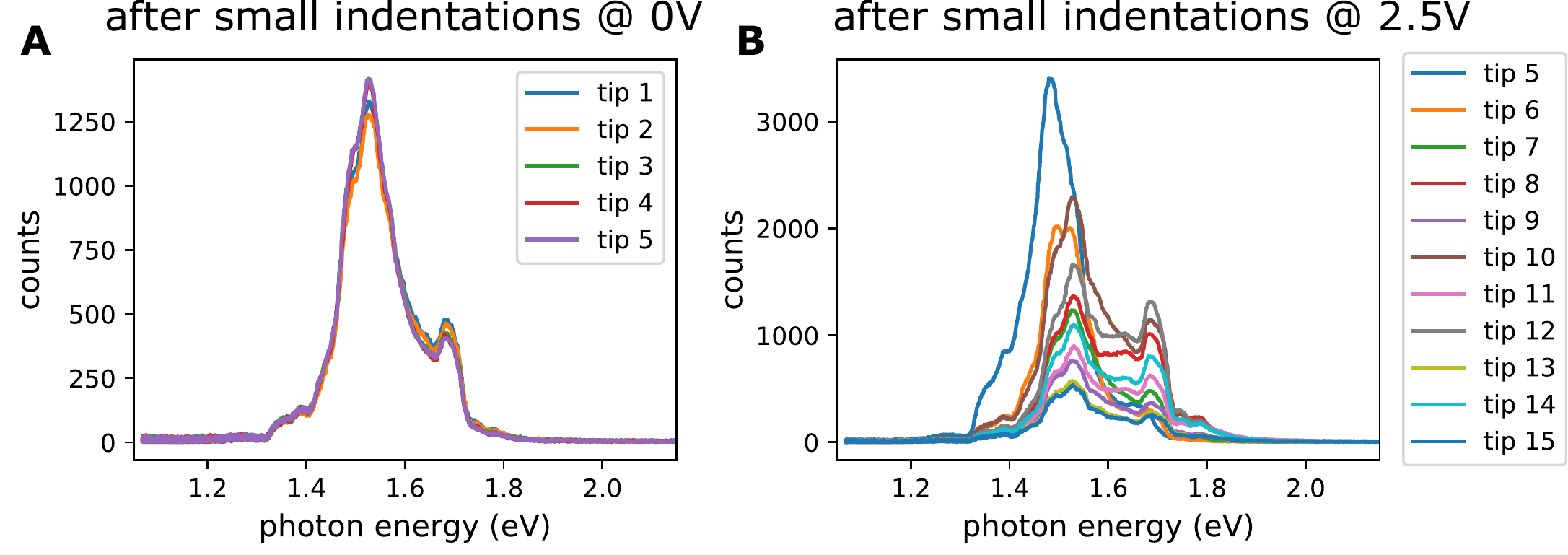}
\caption{\label{fig:SmallPokes}
\textbf{STML spectrum on Au(111) after small tip pokes.} \textbf{A} STML spectra taken on Au(111) after consecutive tip formings by small ($<1\,nm$) pokes into the Au surface at \unit{0}{V}. The emission spectrum is barely changed. \textbf{B} STML spectrum on Au(111) after consecutive small tip formings at \unit{2.5}{V}. The STML is gradually modified in intensity and spectral weight.
}
\end{figure*}

\begin{figure*}[]
\includegraphics[width=\textwidth]{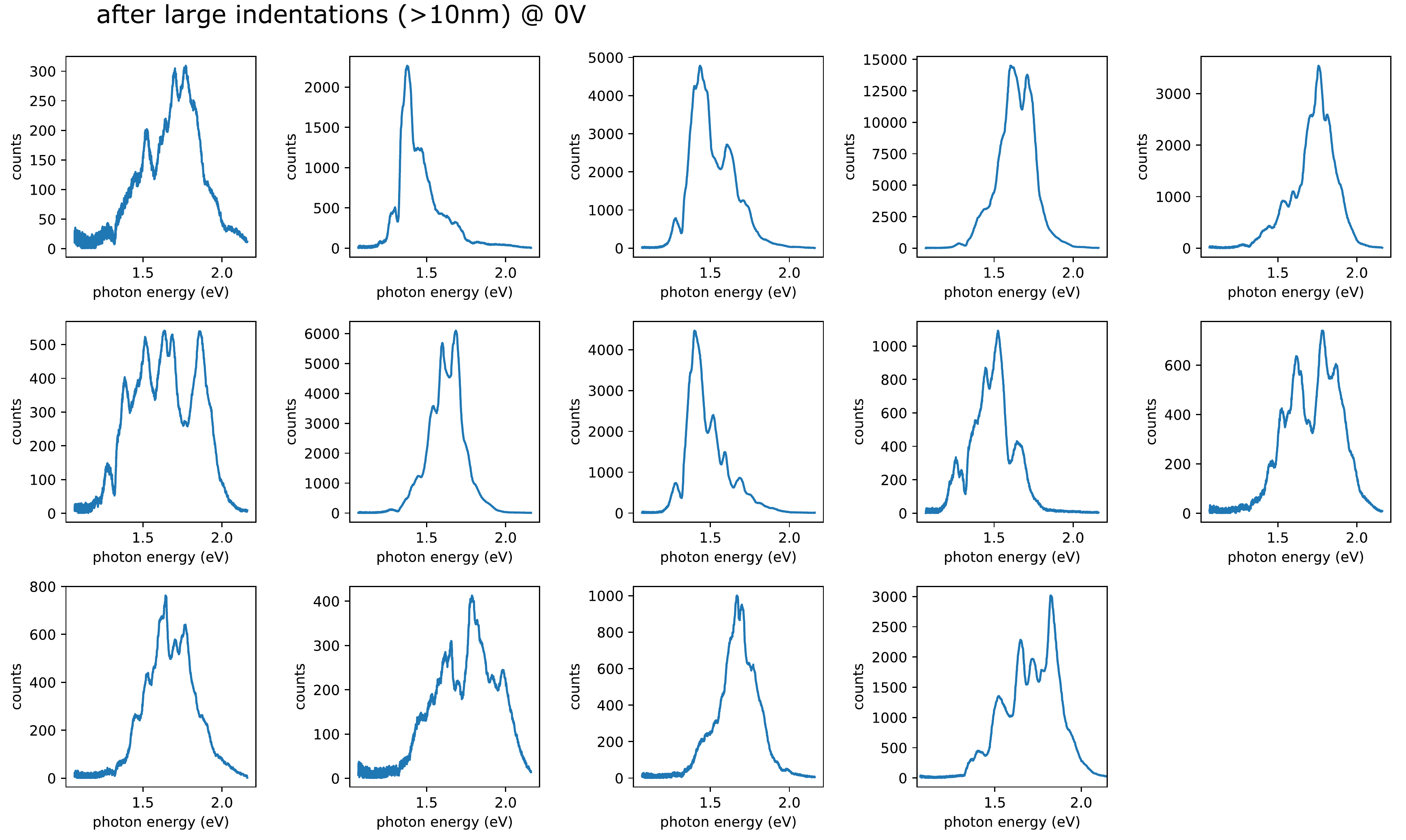}
\caption{\label{fig:BigPokes}
\textbf{STML spectrum on Au(111) after big tip pokes.} STML spectra taken on Au(111) after consecutive tip formings by large ($>10$\,nm) pokes into the Au surface at \unit{0}{V}. The emission spectrum is significantly changed after each poke in both intensity and spectral shape.
}
\end{figure*}

The shape of the tip, however, has a decisive effect on the STML emission spectrum. While after small pokes ($\approx$\unit{1}{nm} approach from the tunneling set-point) at zero bias the STML spectrum is only marginally changed (\figref{fig:SmallPokes}A), the spectrum changes shape when \unit{2.5}{V} are applied during the pokes (\figref{fig:SmallPokes}B). The emission intensity can be dramatically changed but also different spectral ranges become enhanced. In \figref{fig:BigPokes} a series of STML spectra are shown after consecutive big pokes ($>1$\,nm approach from the tunneling set-point) at zero bias. The spectrum is considerably modified after each poke, both in intensity and spectral shape. This shows that the spectral transfer function that modulates the STM luminescence is dominated by the mesoscopic tip shape. It also hints at the potential for spectral enhancement by tailoring the nanocavity formed by the tip and substrate.

\bibliographystyle{science}
\bibliography{STML,refs}